\documentclass[preprint,prc,aps,tightenlines,floats,superscriptaddress,showpacs,showkeys,nofootinbib]{revtex4-1}
\usepackage{graphicx}
\usepackage{dcolumn}
\usepackage{bm}
\usepackage{amsmath}
\usepackage{float}
\usepackage{multirow}
\usepackage{slashed}
\usepackage{xcolor}
\usepackage{physics}
\usepackage{gensymb}
\usepackage{mathtools}
\usepackage{braket}
\usepackage{lipsum}
\usepackage{amsfonts}
\usepackage{amssymb}
\usepackage{epsfig}
\usepackage{ulem,tensor}
\usepackage{tabularx}
\usepackage{epstopdf}

\usepackage[colorlinks=true,pdfstartview=FitV,bookmarks=true,bookmarksnumbered=true,breaklinks]{hyperref}
\usepackage{color}

\begin{document}

\title{Dynamical study of $\pi N\to \pi\pi N$ reactions revisited}
\author{H. Kamano}
\affiliation{Research Center for Nuclear Physics, The University of Osaka, Ibaraki, Osaka 567-0047, Japan}
\author{T.-S. H. Lee}
\email{tshlee@anl.gov}
\affiliation{Physics Division, Argonne National Laboratory, Argonne, Illinois 60439, USA}

\begin{abstract}
Using the Argonne National Laboratory-The University of Osaka (ANL-Osaka) DCC model 
of meson-nucleon reactions, we extend the study of
Phys.~Rev.~C \textbf{79},~025206~(2009) and
Phys.~Rev.~C \textbf{88},~045203~(2013) to predict the cross sections of
the $\pi N \to \pi \pi N$ reactions. 
The model was constructed by fitting 
only the two-body reactions:
$\pi N,\gamma N \to \pi N, \eta N, K\Lambda, K\Sigma$.
Thus, the results for $\pi N \to \pi \pi N$ presented here 
are predictions of the ANL-Osaka DCC model,
which serve to examine the extent to which the forthcoming data from J-PARC can be described. 
This study provides information for improving the extraction of nucleon resonances 
that have large decay widths to $\pi\pi N$ states.
We present results for the total cross sections,
invariant mass distributions, and angular distributions.
We also identify the observables and energy regions 
where the higher mass nucleon resonances in the $S_{31}$, $P_{33}$, $D_{33}$, $F_{37}$,
$D_{13}$, $D_{15}$, and $F_{15}$ partial waves can be most effectively investigated.
\end{abstract}
%-------------------------------------------------------------------------------------

%11.80.La   Multiple scattering
%13.30.Eg   Decays of baryons; Hadronic decays
%13.60.Le   Photon and charged-lepton interactions with hadrons; Meson production
%13.75.Gx   Pion-baryon interactions
%13.88.+e   Polarization in interactions and scattering
%14.20.Gk   Properties of specific particles, Baryon resonances (S=C=B=0)

\maketitle

\section{\label{sec:intro}Introduction}
Recent studies employing coupled-channel approaches have greatly 
advanced our understanding of nucleon resonances 
($N^*$)\footnote{Unless otherwise specified, $N^*$ refers to 
isospin $I=1/2$ nucleon resonances and $I=3/2$ $\Delta$ resonances.}
(see, e.g., Ref.~\cite{Doring:2025sgb}).
However, significant discrepancies among the results 
reported by various analysis groups remain to be resolved. 
These discrepancies are primarily observed for $N^*$ states in the energy region where
two-pion production dominates the $\pi N$ reaction cross sections.
For such $N^*$ states, coupled-channel effects arising from three-body $\pi\pi N$ scattering states 
are expected to have a significant impact on their properties. 
Therefore, a comprehensive coupled-channel analysis that incorporates two-pion production data is 
indispensable for determining the properties of these $N^*$ states.

Despite the importance of two-pion production data for firmly establishing $N^*$ states,
high-statistics data for the $\pi N \to \pi \pi N$ reaction suitable for 
detailed partial-wave analysis remain limited.
Notable examples of precise measurements include the $\pi^- p \to \pi^0 \pi^0 n$ data 
obtained by Crystal Ball in 2004~\cite{crystalball},
and more recently, the $\pi^- p \to \pi^- \pi^0 p$ and $\pi^- p \to \pi^+ \pi^- p$ measurements by HADES in 2020~\cite{hades}.
These data have been incorporated in a recent partial-wave analysis~\cite{boga}. 
However, these experiments were limited to relatively low energies ($W \lesssim 1.55$~GeV).
Therefore, it is highly desirable to obtain such high-statistics data over the wide energy range up to $W \sim 2$~GeV.
In this regards, the upcoming experiment at J-PARC~\cite{j-parc} is expected to address this need.
In this experiment, it is planned to measure the four reaction channels:
$\pi^+ p \to \pi^+ \pi^+ n$, $\pi^+ p \to \pi^+ \pi^0 p$, 
$\pi^- p \to \pi^+ \pi^- n$, and $\pi^- p \to \pi^- \pi^0 p$.
Also, the beam momenta used in the first measurement is planned to be $0.98$--$1.15$~GeV/c, 
which corresponds to $W = 1.66$--$1.76$~GeV. 
The J-PARC experiment aims at both of the $\pi^+ p$ and $\pi^- p$ reactions, and 
covers a higher energy region than those of Crystal Ball and HADES.

Given the importance of two-pion production processes in establishing $N^*$ states,
we have investigated the $\pi N \to \pi \pi N$ reactions in a series of works
based on a dynamical coupled-channel (DCC) model, 
the Argonne National Laboratory-The University of Osaka (ANL-Osaka) DCC 
model~\cite{ao96,ao07-1,ao07-2,ao09-1,ao09-2,ao09-3,ao10,ao13-1,ao13-2,ao15-1,ao16,ao18,ao19,ao14,ao15-2}.
Our first investigation of the $\pi N \to \pi \pi N$ reactions~\cite{ao09-1} 
employed the DCC model developed in Ref.~\cite{ao07-2}.
This model incorporated five reaction channels ($\pi N$, $\eta N$, and 
the three quasi-two-body channels of $\pi \pi N$: $\pi\Delta$, $\rho N$, and $\sigma N$),
and the model parameters were determined by fitting to the SAID amplitudes~\cite{said}
of $\pi N$ elastic scattering up to $W = 2$~GeV.
The model predicted the total cross sections and invariant mass distributions for 
$\pi N \to \pi \pi N$, describing the existing data reasonably well 
over a wide energy range from the threshold up to $W = 2$ GeV.
It clearly demonstrated the critical role of unitarized
coupled-channel effects in comprehensively describing $\pi N$
reactions with various final states.
In Ref.~\cite{ao13-2}, we further investigate the $\pi N \to \pi \pi N$ reactions 
using the DCC model published in 2013~\cite{ao13-1}.
The 2013 model incorporated seven reaction channels 
($\pi N$, $\eta N$, $K\Lambda$, $K\Sigma$, $\pi\Delta$, $\rho N$, and $\sigma N$),
and the model parameters were determined by fitting to the existing data for 
$\pi N \to \pi N, \eta N, K\Lambda, K\Sigma$ and 
$\gamma p \to \pi N, \eta N, K\Lambda, K\Sigma$ up to $W = 2$ GeV.
This investigation particularly focused on clarifying the ambiguities in
the $N^*$ resonance parameters that arise when performing 
a comprehensive coupled-channel analysis excluding
the $\pi N \to \pi \pi N$ data.
The examination presented in Sec.~IV of Ref.~\cite{ao13-2}
showed that the coupled-channel analysis without the
$\pi N \to \pi \pi N$ data indeed leaves sizable ambiguities in 
the $N^*$ parameters, particularly those associated with 
the three-body $\pi \pi N$ channel. 
This indicates that the $\pi N \to \pi \pi N$ data 
are crucial for resolving these ambiguities.

Motivated by the activities of recent and upcoming experiments regarding 
the $\pi N \to \pi \pi N$ reactions mentioned above, in this work we
extend our previous studies~\cite{ao09-1,ao13-2} to further investigate 
the potential impact of $\pi N \to \pi \pi N$ data on
the determination of $N^*$ parameters.
As a primary objective, we examine the sensitivity
of the $N^*$ states to the $\pi N \to \pi \pi N$ cross sections,
thereby providing useful information for the upcoming experiment at J-PARC.
For this purpose, we employ the ANL-Osaka DCC model published in 2016~\cite{ao16}
(hereafter referred to as the 2016 model).
This model extends the 2013 model by including $\gamma n \to \pi N$ data 
in the fit, and provides a superior description of the data
for $\pi N \to \pi N, \eta N, K\Lambda, K\Sigma$ and 
$\gamma p \to \pi N, \eta N, K\Lambda, K\Sigma$.

In principle, one could also utilize $\gamma^{(*)} N \to \pi \pi N$ data 
to investigate two-pion production off the nucleon.
However, describing photoproduction processes introduces additional model 
parameters associated with the electromagnetic interaction.
Some of these parameters are sensitive specifically to 
the $\gamma N \to \pi \pi N$ reaction and are difficult to determine using 
only single-meson photoproduction data.
Furthermore, from a technical standpoint, the computation of 
$\gamma N \to \pi \pi N$ observables is computationally quite demanding within DCC approaches.
Therefore, data from purely hadronic $\pi N \to \pi \pi N$ processes 
are highly desirable as a first step to clarify the effects of reaction dynamics 
associated with the three-body $\pi\pi N$ channel on $N^*$ states.

It is useful to briefly review the ANL-Osaka DCC model.
Here we focus on hadronic reactions.
This model is based on the meson-exchange mechanisms
and the assumption that the (bare) $N^*$ states emerge
through the $\alpha \to N^*$ vertex interactions where $\alpha$ is 
the meson-baryon states with strangeness $S=0$.
Within the Hamiltonian formulation~\cite{ao96,ao07-1}, 
the unitary condition requires that the partial-wave amplitudes 
$T_{\beta,\alpha}(p_{\beta},p_{\alpha};W)$, which are specified by 
the total angular momentum $J$, parity $P$, and total isospin $I$ 
(these indices are suppressed here), 
are given by the following coupled-channel equations:
\begin{equation}
T_{\beta,\alpha}(p_\beta,p_\alpha;W)
= 
  V_{\beta,\alpha}(p_\beta,p_\alpha;W)
+ \sum_{\gamma} \int p^2 dp V_{\beta,\gamma}(p_\beta,p;W)
  G_\gamma(p;W) T_{\gamma,\alpha}(p,p_\alpha;W),
\label{eq:cc-eq}
\end{equation}
with
\begin{equation}
V_{\beta,\alpha}(p_\beta,p_\alpha;W)
= 
  v_{\beta,\alpha}(p_\beta,p_\alpha)
+ Z_{\beta,\alpha}(p_\beta,p_\alpha; W)
+ \sum_{N^*} \frac{\Gamma^\dagger_{N^*,\beta}(p_\beta)\Gamma_{N^*,\alpha}(p_\alpha)}
                  {W-M^0_{N^*}},
\label{eq:cc-v}
\end{equation}
where $\alpha,\beta,\gamma=\pi N, \eta N, \pi\Delta, \sigma N, \rho N, K\Lambda, K\Sigma$;
$G_\gamma(p;W)$ is the Green's function of the channel $\gamma$;
$M^0_{N^*}$ is the mass of the bare $N^*$ state; 
$v_{\beta,\alpha}(p_\beta,p_\alpha)$ represents
meson-exchange mechanisms derived from effective Lagrangians;
the energy-dependent $Z_{\beta,\alpha}(p_\beta,p_\alpha;W)$ term 
is the one-particle potential that produces the three-body $\pi\pi N$ cut; and
the vertex interaction $\Gamma_{N^*,\alpha}(p_\alpha)$ defines 
the $\alpha \to N^*$ transition.
The model parameters contained in $V_{\beta,\alpha}(p_\beta,p_\alpha;W)$
are determined by fitting to the world data of $\pi N$ and $\gamma^{(*)} N$ reactions,
and the $N^*$ parameters such as pole masses and residues of the amplitudes at the pole
are then extracted by making the analytic continuation of 
$T_{\beta,\alpha}(p_\beta,p_\alpha;W)$ to the complex-energy plane.

This paper is organized as follows.
In Sec.~\ref{sec:formulation}, we present the formulas for 
calculating  the cross sections of $\pi N \to \pi\pi N$ reactions
within the ANL-Osaka DCC model.
The results and discussions are given in Sec.~\ref{sec:results}. 
The summary is given in Sec.~\ref{sec:summary}.

\section{\label{sec:formulation}Formulation}
In this section, we present the formulas for calculating the cross sections
for the $\pi N \to \pi \pi N$ reactions used in this paper.

\subsection{Kinematics and cross sections}
In the total center-of-mass (CM) frame, the momentum variables for
the $2\to 3$ reactions can be specified as
\begin{equation}
a(\vec{p}_a)+ b(\vec{p}_b)
\to
c(\vec{p}_c) + d(\vec{p}_d) + e(\vec{p}_e),
\label{eq:process}
\end{equation}
where $\vec{p}_a + \vec{p}_b= \vec{p}_c + \vec{p}_d +\vec{p}_e=0$.
For later use, we introduce $\vec{k}$ and $\vec{k'}$ defined by
$\vec{k} = \vec{p}_a = -\vec{p}_b$ and $\vec{k'} = \vec{p}_c + \vec{p}_d = -\vec{p}_e$, respectively.
The magnitude of $\vec{k}$ (denoted as $k$) is determined by the relation $W=E_a(k) + E_b(k)$,
where $W$ is the total scattering energy in the total CM frame
and $E_a(k) = \sqrt{m_a^2+k^2}$ is the relativistic energy for a particle $a$ with mass $m_a$ and momentum $\vec{k}$.
Similarly, the magnitude of $\vec{k'}$ (denoted as $k'$) is given by $W=E_e(k') + E_{cd}(k')$,
where $E_{cd}(k')$ is defined by $E_{cd}(k') = \sqrt{M_{cd}^2 + k'^2}$ 
with $M_{cd}$ being the invariant mass of the $c$-$d$ pair.
In the case of $\pi N \to \pi \pi N$ reaction, $(c,d,e)$ in the final state 
corresponds to any allowed charge states formed from two pions and one nucleon.

Following the convention of Goldberger and Watson~\cite{gw},
the total cross sections and unpolarized differential cross sections 
investigated in this work are given by
\begin{equation}
\sigma = 
\int^{W-m_e}_{m_c+m_d} dM_{cd} \frac{d\sigma}{dM_{cd}} = 
\int d\Omega_{k'} \frac{d\sigma}{d\Omega_{k'}},
\label{eq:tcrs}
\end{equation}
\begin{equation}
\frac{d\sigma}{dM_{cd}} = \int d\Omega_{k'} \frac{d\sigma}{dM_{cd}d\Omega_{k'}},
\label{eq:dsdm}
\end{equation}
\begin{equation}
\frac{d\sigma}{d\Omega_{k'}} = \int^{W-m_e}_{m_c+m_d} dM_{cd} \frac{d\sigma}{dM_{cd}d\Omega_{k'}},
\label{eq:dsdx}
\end{equation}
with
\begin{equation}
\frac{d\sigma}{dM_{cd}d\Omega_{k'}} =
16\pi^3 \frac{\rho_{ab}}{k^2}
\int d\Omega_{k_{cd}} 
\frac{k_{cd}k'}{W} E_c(p_c)E_d(p_d)E_e(p_e)
\overline{\sum_{\textrm{spins}}} |T|^2,
\label{eq:dsdmdo}
\end{equation}
where $\Omega_p$ denotes the solid angle for the momentum $\vec p$; 
$\rho_{ab} = \pi kE_a(k)E_b(k)/W$; 
$k_{cd}$ is the magnitude of the relative momentum between $c$ and $d$
in the CM frame of the $c$-$d$ pair, 
which is given by $M_{cd} = E_c(k_{cd}) + E_d(k_{cd})$;
$T$ is the $T$-matrix element for the reaction;
and the symbol $\overline{\sum}_{\textrm{spins}}$ in Eq.~(\ref{eq:dsdmdo}) 
denotes a summation over all final spins and an average over initial spins.

\subsection{$T$-matrix element for $\pi N \to \pi \pi N$ in the ANL-Osaka DCC model}
\begin{figure}[t]
\begin{center}
\includegraphics[width=0.8\textwidth,angle=0,clip]{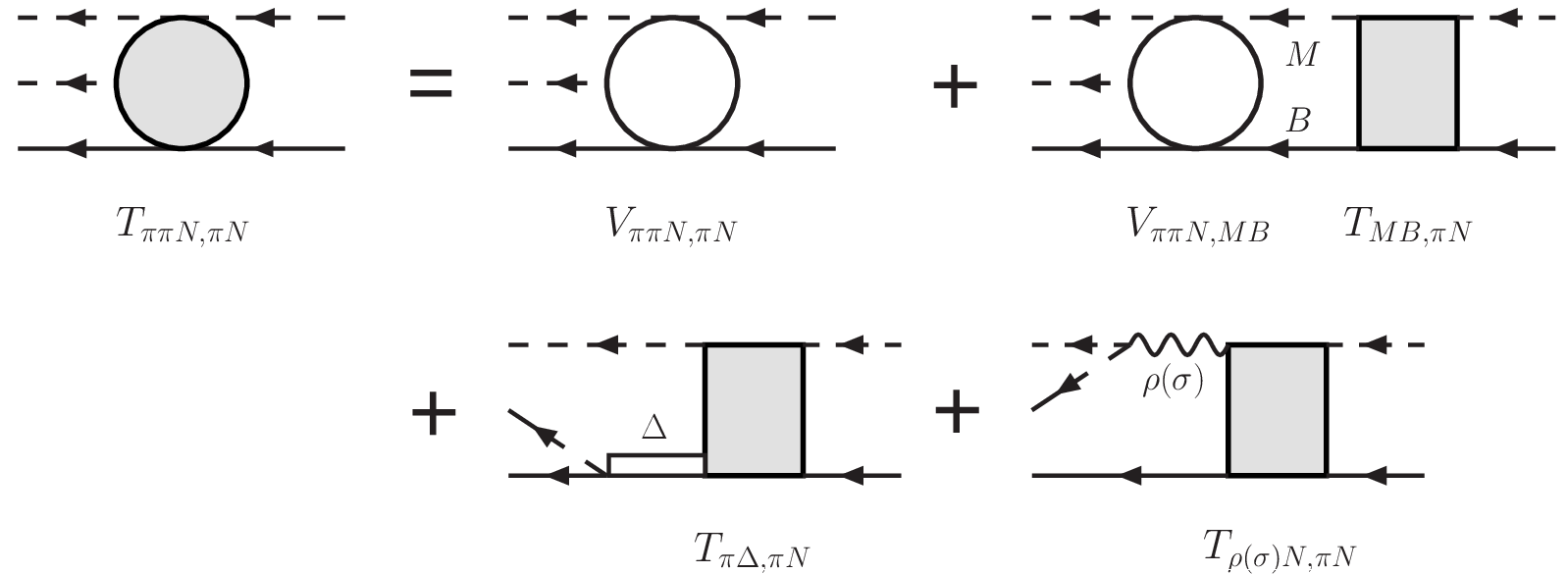}
\caption{Schematic representation of the $\pi N \to \pi\pi N$ reaction mechanisms. The figure is from Ref.~\cite{ao13-2}.}
\label{fig:diagram}
\end{center}
\end{figure}
In the framework of the ANL-Osaka DCC model~\cite{ao07-1,ao09-1}, 
the $\pi N\to\pi \pi N$ reaction is described by the diagrams depicted in Fig.~\ref{fig:diagram}.
The full $T$-matrix element in Eq.~(\ref{eq:dsdmdo}) is expressed as:
\begin{equation}
T
=
T^{\text{dir}}
+\sum_{MB=\pi\Delta,\sigma N, \rho N}T^{MB}, 
\label{eq:amp}
\end{equation}
with
\begin{eqnarray}
T^{\text{dir}}
&=&
 V_{\pi \pi N, \pi N}+
\sum_{MB} 
V_{\pi\pi N,MB} G_{MB} T_{MB,\pi N} ,
\label{eq:tdir}
\\
T^{\pi\Delta}
&=&
\Gamma_{\pi N, \Delta} G_{\pi \Delta} T_{\pi\Delta, \pi N} ,
\\
T^{\rho N}
&=&
\Gamma_{\pi \pi, \rho} G_{\rho N} T_{\rho N, \pi N} ,
\\
T^{\sigma N}
&=&
\Gamma_{\pi \pi, \sigma} G_{\sigma N} T_{\sigma N, \pi N} .
\end{eqnarray}
In these expressions, $V_{\pi \pi N, MB}$ represents the potential 
for the direct transition from two-body to three-body states~\cite{ao09-1},
while $G_{MB}$ denotes the Green's function for the channel $MB$.
The factors $\Gamma_{\pi N,\Delta}$, $\Gamma_{\pi\pi,\rho}$, and $\Gamma_{\pi\pi,\sigma}$ 
correspond to the decay vertices for 
$\Delta\to\pi N$, $\rho \to \pi \pi$, and $\sigma \to \pi \pi$, respectively.
The summation over $MB$ in Eq.~(\ref{eq:tdir}) includes the channels 
$MB=\pi N, \eta N, \pi \Delta, \rho N, \sigma N, K\Lambda, K\Sigma$.
We compute the two-body plane-wave amplitudes, $T_{MB, \pi N}$, 
from the partial-wave amplitudes derived by solving Eq.~(\ref{eq:cc-eq}).
For full details regarding the two-body amplitudes, meson-baryon Green's functions, 
and decay vertices, we refer the reader to Ref.~\cite{ao07-1,ao09-1}.
Concerning the direct term $T^{\textrm{dir}}_{\pi\pi N,\pi N}$, however, 
we implement the practical prescription 
given in Eq.~(16) of Ref.~\cite{ao09-1}, 
which simplifies the amplitude to:
\begin{equation}
T^{\text{dir}}_{\pi \pi N, \pi N}
\sim
 \tilde V_{\pi \pi N, \pi N} +
\Gamma_{\pi N, N} G_{\pi N} T_{\pi N, \pi N}.
\label{eq:tdir2}
\end{equation}
Here, $\tilde V_{\pi \pi N, \pi N}$ denotes the direct $\pi N \to \pi \pi N$ transition potential 
associated with diagrams (f)-(k) in Fig.~2 of Ref.~\cite{ao09-1}, 
and $\Gamma_{\pi N, N}$ is the $N\to \pi N$ vertex function.

\section{\label{sec:results}Results and discussions}

\subsection{\label{sec:tcrs}Total cross sections}
\begin{figure}[t]
\begin{center}
\includegraphics[width=1.0\textwidth,angle=0,clip]{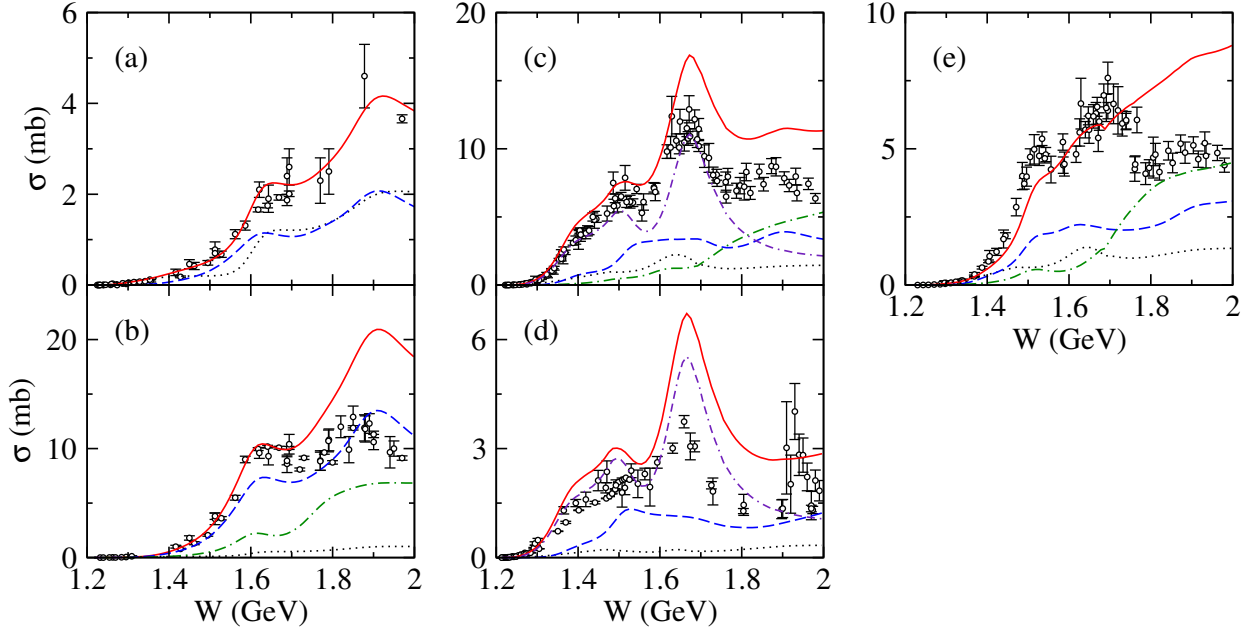}
\caption{
Total cross sections of 
(a)~$\pi^+ p \to \pi^+\pi^+ n$, (b)~$\pi^+p \to \pi^+\pi^0p$, 
(c)~$\pi^- p \to \pi^+\pi^- n$, (d)~$\pi^-p \to \pi^0\pi^0 n$, and
(e)~$\pi^- p \to \pi^-\pi^0 p$.
Solid (red) curves are the full results; 
dashed (blue) curves are the results of $T^{\pi\Delta}$ process only; 
dash-dot (green) curves are the results of $T^{\rho N}$ process only; 
dash-dash-dot (indigo) curves are the results of $T^{\sigma N}$ process only; 
and dotted (black) curves are the results of $T^\textrm{dir}$ process only.
See Refs.~\cite{manley,ao09-1} and references therein for the data.
}
\label{fig:tcrs}
\end{center}
\end{figure}
Figure~\ref{fig:tcrs} shows the total cross sections for the $\pi N \to \pi \pi N$ 
reactions based on the predictions obtained with the 2016 model~\cite{ao16}.
The predicted full result for the $\pi^+ p \to \pi^+ \pi^+ n$ reaction 
is found to be in good agreement with the existing data up to $W=2$~GeV.
For the other four reactions,
our results show reasonable agreement with the data at low energies;
however, they significantly overestimate the data above $W \sim 1.65$~GeV.
This behavior is consistent with our previous results reported in 2013~\cite{ao13-2}.
In the same figure, we also show the contributions of each reaction process 
defined in Eq.~(\ref{eq:amp}) to the total cross sections.
For the $\pi^+ p \to \pi^+ \pi^+ n$ reaction [Fig.~\ref{fig:tcrs}(a)], 
only the isospin $I=3/2$ components of $T^{\pi\Delta}$ and $T^\textrm{dir}$ 
contribute, and these contributions are comparable over the entire energy range up to $W=2$~GeV.
On the other hand, contributions arise from the isospin $I=3/2$ components of 
$T^{\pi\Delta}$, $T^{\rho N}$, and $T^{\textrm{dir}}$
in the $\pi^+ p \to \pi^+ \pi^0 p$ reaction [Fig.~\ref{fig:tcrs}(b)]. 
Given that the $T^{\textrm{dir}}$ process gives 
a negligible contribution compared to the other two processes for this reaction,
fine-tuning of the isospin $I=3/2$ components of $T^{\pi\Delta}$ and $T^{\rho N}$
will be required to resolve the significant overestimation of 
the total cross sections above $W \sim 1.7$~GeV.
For the $\pi^- p \to \pi^+ \pi^- n$ and 
$\pi^- p \to \pi^0 \pi^0 n$ reactions
[Fig.~\ref{fig:tcrs}(c) and (d)], 
the $T^{\sigma N}$ process dominates the cross sections at low energies
and is responsible for the peak at $W \sim 1.65$~GeV.
Contributions from the $T^{\pi\Delta}$ and $T^{\rho N}$ processes 
are also non-negligible and become comparable to those of $T^{\sigma N}$ above $W\sim 1.8$~GeV,
while the $T^{\textrm{dir}}$ process provides a minor contribution, particularly for
$\pi^- p \to \pi^0 \pi^0 n$.
The consistent overestimation of these cross sections up to $W=2$~GeV
suggests that the decays of $I=1/2$ $N^*$ states to 
$\pi\Delta$, $\rho N$, and $\sigma N$ in particular
are poorly constrained by the partial-wave analysis of single-meson production alone.
Consequently, incorporating $\pi N \to \pi \pi N$ data into the analysis is essential.
For the $\pi^- p \to \pi^- \pi^0 p$ channel [Fig.~\ref{fig:tcrs}(e)],
the contributions from $T^{\pi\Delta}$, $T^{\rho N}$ and $T^{\textrm{dir}}$ 
are of comparable magnitude throughout the energy range up to $W=2$~GeV.
Resolving the overestimation above $W\sim 1.8$~GeV
necessitates refinements to all three processes.
We also note that a small cusp structure appears at $W\sim 1.68$~GeV
in the $\pi^- p \to \pi^- \pi^0 p$ total cross section.
This behavior is attributed to the opening of the $K\Sigma$ channel
and arises from the $S$-wave $\pi N \to \rho N$ amplitude in $T^{\rho N}$.

\begin{figure}[t]
\begin{center}
\includegraphics[width=0.75\textwidth,angle=0,clip]{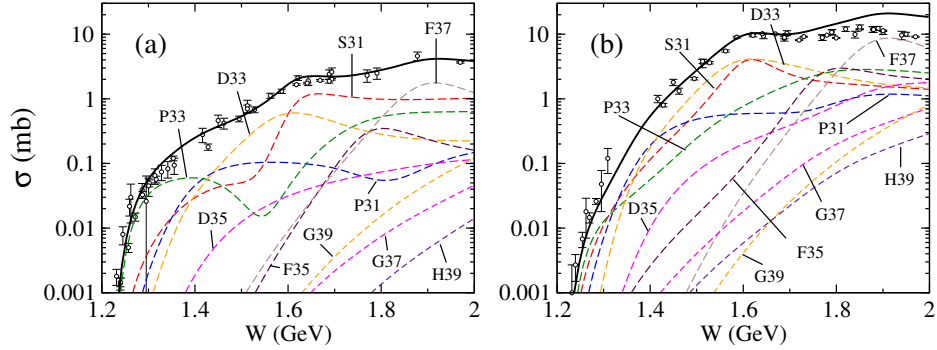}
\caption{
Contribution of each partial wave to the $\pi^+ p \to \pi \pi N$ total cross sections:
(a)~$\pi^+ p \to \pi^+ \pi^+ n$ and (b)~$\pi^+ p \to \pi^+ \pi^0 p$.
Solid curves are the full results, while dashed curves are contributions of each partial wave
to the cross sections.
See Refs.~\cite{manley,ao09-1} and references therein for the data.
}
\label{fig:tcrs-pp-pwa}
\end{center}
\end{figure}
\begin{figure}[t]
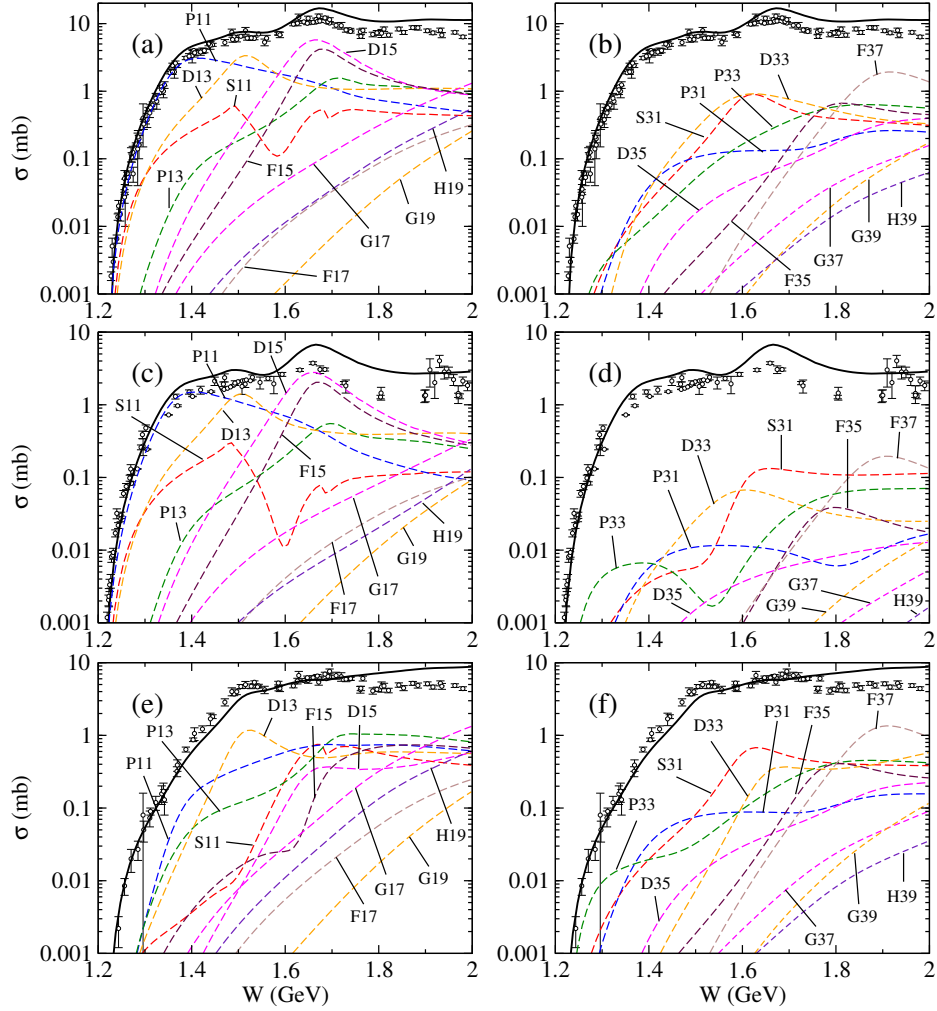

\begin{center}
\includegraphics[width=0.75\textwidth,angle=0,clip]{tcrs-pwa-pmn.eps}
\includegraphics[width=0.75\textwidth,angle=0,clip]{tcrs-pwa-00n.eps}
\includegraphics[width=0.75\textwidth,angle=0,clip]{tcrs-pwa-0mp.eps}
\caption{
Contribution of each partial wave to the $\pi^- p \to \pi \pi N$ total cross sections.
Panels~(a) and (b)~$\pi^- p \to \pi^+ \pi^- n$, 
(c) and (d)~$\pi^- p \to \pi^0 \pi^0 n$, and
(e) and (f)~$\pi^- p \to \pi^-\pi^0 p$.
Panels (a), (c), and (e) [(b), (d), and (f)] present $I=1/2$ [$I=/3/2$] partial waves.
Solid curves are the full results, while dashed curves are contributions of each partial wave
to the cross sections.
See Refs.~\cite{manley,ao09-1} and references therein for the data.
}
\label{fig:tcrs-mp-pwa}
\end{center}
\end{figure}
Figures~\ref{fig:tcrs-pp-pwa} and~\ref{fig:tcrs-mp-pwa} show
the partial-wave contributions to the total cross sections up to $J=9/2$.
Generally, the behavior of each partial wave is consistent with our previous results 
based on the 2013 model~\cite{ao13-1}.
Regarding the initial $\pi^+ p$ reactions [Fig.~\ref{fig:tcrs-pp-pwa}],
the $P_{33}$ partial wave dominates the $\pi^+ p \to \pi^+ \pi^+ n$ cross section
from the threshold to $W=1.4$~GeV, while 
several partial waves contribute comparably to 
$\pi^+ p \to \pi^+ \pi^0 n$ at low energies.
In the $W=1.5$--$1.75$~GeV region, however,
$S_{31}$ and $D_{33}$ dominate both reactions,
with $F_{37}$ taking over above $W=1.8$~GeV.
For the initial $\pi^- p$ reactions [Fig.~\ref{fig:tcrs-mp-pwa}], 
$I=3/2$ partial waves represent sub-dominant contributions,
except in the $\pi^+\pi^-n$ and $\pi^-\pi^0p$ final states at $W\sim 1.9$ GeV,
where $F_{37}$ is comparable to the largest $I=1/2$ waves.
Between $W=1.5$ and~$1.75$~GeV,
$D_{13}$ and $D_{15}$ are dominant in all three charged states,
as is $F_{15}$ for the $\pi^+\pi^- n$ and $\pi^0\pi^0 n$ final states.
We emphasize that the $P_{11}$ partial wave dominates 
the $\pi^- p \to \pi^+\pi^- n$ and $\pi^- p \to \pi^0\pi^0 n$ 
up to $W\sim 1.4$ GeV, corresponding to the Roper resonance region.
As mentioned in Ref.~\cite{ao13-2},
this contrasts with photoproduction,
where the Roper resonance makes a minor contribution and
is overshadowed by the prominent first $D_{13}$ resonance.
Thus $\pi N \to \pi \pi N$ reactions offer crucial insights
into high-mass $N^\ast$ states and the elusive Roper resonance.

\subsection{\label{sec:dcrs}Invariant mass distributions and angular distributions}
In this subsection, we present our predictions for invariant mass distributions
and angular distributions from the 2016 model.
First we focus on the low energy region below $W=1.55$~GeV,
where the precise data for $\pi^- p \to \pi \pi N$ from 
Crystal Ball and HADES are available.

\begin{figure}[t]
\begin{center}
\includegraphics[width=0.6\textwidth,angle=0,clip]{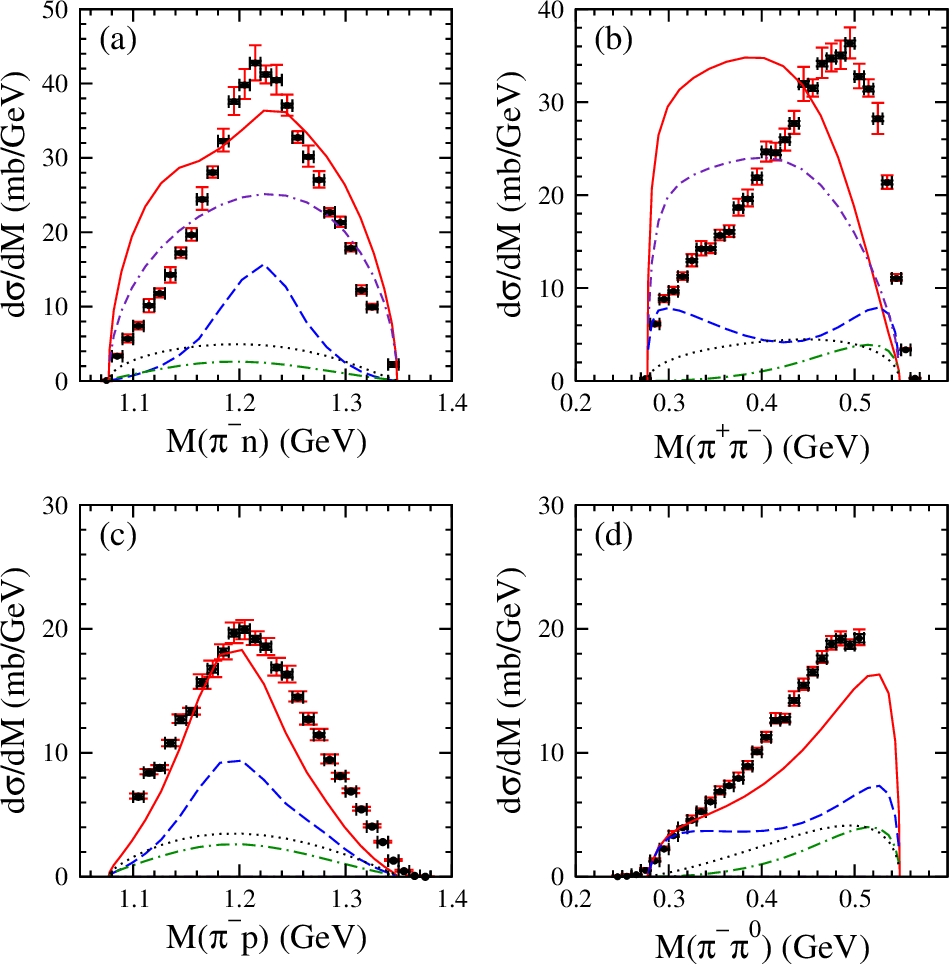}
\caption{
Comparison with the HADES data~\cite{hades} for the invariant mass distributions at
$W=1487$~MeV ($p_{\textrm{in}}=685$~MeV/c with $p_{\textrm{in}}$ being pion beam momenta).
Panels~(a) and~(b) are of $\pi^- p \to \pi^+\pi^- n$, and
panels~(c) and~(d) are of $\pi^- p \to \pi^-\pi^0 p$.
The meaning of each curve is the same as in Fig.~\ref{fig:tcrs}
}
\label{fig:dsdm-hades}
\end{center}
\end{figure}
\begin{figure}[t]
\begin{center}
\includegraphics[width=0.6\textwidth,angle=0,clip]{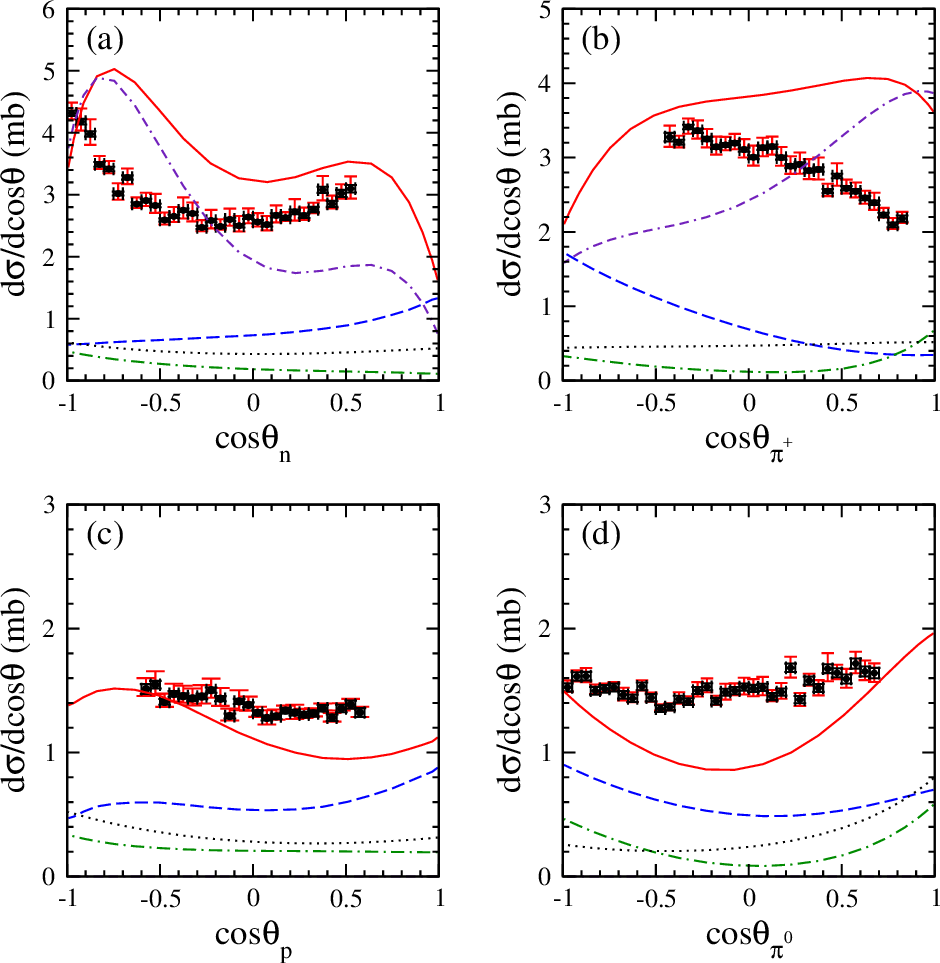}
\caption{
Comparison with the HADES data~\cite{hades} for the angular distributions at
$W=1487$~MeV ($p_{\textrm{in}}=685$~MeV/c with $p_{\textrm{in}}$ being pion beam momenta).
Panels~(a) and~(b) are of $\pi^- p \to \pi^+\pi^- n$, and
panels~(c) and~(d) are of $\pi^- p \to \pi^-\pi^0 p$.
The meaning of each curve is the same as in Fig.~\ref{fig:tcrs}.
}
\label{fig:dsdx-hades}
\end{center}
\end{figure}
Figures~\ref{fig:dsdm-hades} and~\ref{fig:dsdx-hades} compare
our predictions with the HADES data at $W=1487$~MeV.
HADES provides the data for the $\pi^- p \to \pi^+\pi^- n$ 
and $\pi^- p \to \pi^-\pi^0 p$ reactions.
We observe that the predicted invariant 
mass distributions for $\pi^- p \to \pi^+\pi^- n$
show significant deviations from the data,
as seen in Fig.~\ref{fig:dsdm-hades}~(a) and~(b).
As demonstrated in Refs.~\cite{kamano1,kamano2}, 
strong interference between $T^{\pi\Delta}$ and $T^{\sigma N}$ is crucial 
for describing the shape of the invariant mass distributions 
for $\pi^- p \to \pi^+\pi^- n$ and $\pi^- p \to \pi^0\pi^0 n$
in the Roper resonance region.
Since the predicted invariant mass distributions are largely governed by
the behavior of the $T^{\sigma N}$ process, 
the observed discrepancy implies that
the $N^* \to \pi\Delta$ decay is underestimated and/or 
the $N^* \to \sigma N$ decay is overestimated in the 2016 model.
This discrepancy is corroborated by
the angular distributions of $\pi^- p \to \pi^+\pi^- n$ 
[Fig.~\ref{fig:dsdx-hades}~(a) and~(b)],
where the predictions exhibit a significantly different angular 
dependence compared to the data.
Thus, the HADES data for $\pi^- p \to \pi^+\pi^- n$ provide
crucial constraints for determining the $\pi\Delta$ 
and $\sigma N$ couplings to $N^*$ in the Roper resonance region.
For the $\pi^- p \to \pi^-\pi^0 p$ reaction,
our predictions reproduce the general trends of the data.
However, discrepancies appear in the high invariant-mass region
of the invariant mass distributions
[Fig.~\ref{fig:dsdm-hades}~(c) and~(d)]
and in the angular distributions at $\cos\theta_p \gtrsim 0$ [Fig.~\ref{fig:dsdx-hades}~(c)]
and around $\cos\theta_{\pi^o} \sim 0$ [Fig.~\ref{fig:dsdx-hades}~(d)].
Although $T^{\pi\Delta}$ is dominant,
the contributions from $T^{\rho N}$ and $T^{\textrm{dir}}$ are non-negligible.
Fine-tuning these three processes will be required to better describe the data.

\begin{figure}[t]
\begin{center}
\includegraphics[width=0.6\textwidth,angle=0,clip]{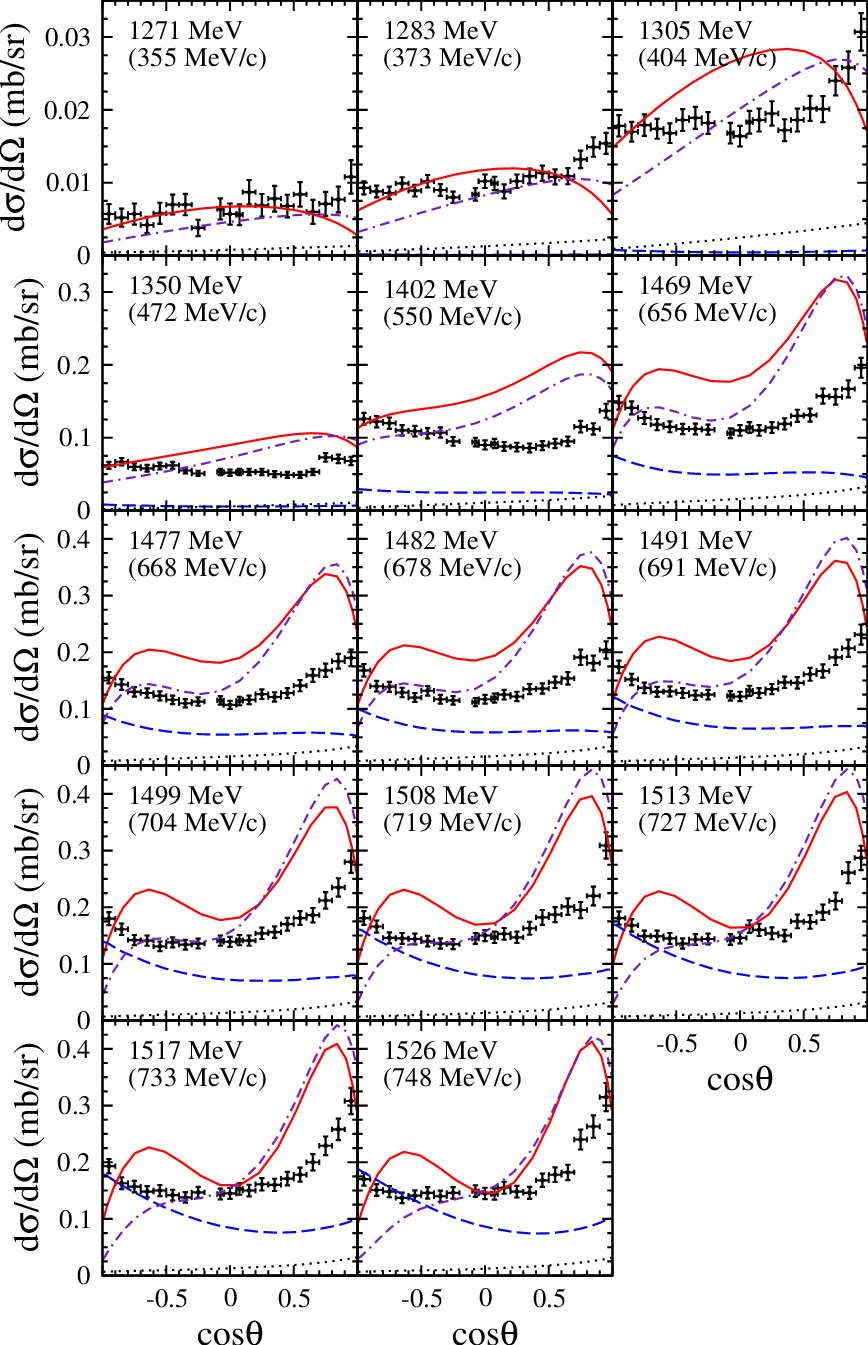}
\caption{
Comparison with the Crystal Ball data~\cite{crystalball} for the angular distributions
of $\pi^- p \to \pi^0 \pi^0 n$.
The solid angle $\Omega$ and scattering angle $\theta$ are of the outgoing $2\pi^0$ system.
In each panel, the corresponding $W$ ($p_{\textrm{in}}$ in parenthesis) is presented.
The meaning of each curve is the same as in Fig.~\ref{fig:tcrs}.
}
\label{fig:dsdO-cb}
\end{center}
\end{figure}
Figure~\ref{fig:dsdO-cb} shows the comparison with the angular distribution data 
for $\pi^- p \to \pi^0 \pi^0 n$ from Crystal Ball, 
covering the energy range from near the threshold to $W\sim 1530$ MeV.
Since neither $T^{\rho N}$ nor other $\pi N \to M N$ processes
(where $M$ is an unflavored $I=1$ meson) contribute to this channel,
this reaction is particularly useful for investigating
the $\sigma N$ coupling to $N^*$.
Similar to the case of $\pi^- p \to \pi^+ \pi^- n$ [Fig.~\ref{fig:dsdx-hades}~(a) and~(b)], 
where $T^{\sigma N}$ also contributes, 
a major source of the significant deviation from the data arises from 
the large contribution of the $T^{\sigma N}$ process. 
Adjusting the relative strength and interference 
between $T^{\sigma N}$ and the other processes will be key to resolving 
the discrepancies between our results and the data at forward and backward angles.

Although the existing data for $\pi^- p \to \pi\pi N$ 
from Crystal Ball and HADES are of particular interest 
in connection with the Roper resonance, 
we need precise differential cross section data 
in the higher energy region as well, 
in order to establish the high-mass $N^*$ resonances 
where the double pion production dominates $\pi N$ reactions.
Furthermore, data for $\pi^+ p \to \pi \pi N$ are highly desirable.
Due to isospin filtering,
the $\pi^+ p$ reactions provide an ideal platform to exclusively investigate 
the isospin $I=3/2$ $\Delta^*$ resonances.
The upcoming experiment at J-PARC offers 
a promising opportunity to address these issues.
It is planned to utilize both $\pi^-$ and $\pi^+$ beams
to measure the reactions $\pi^- p \to \pi^+\pi^- n$, $\pi^- p \to \pi^-\pi^0 p$,
$\pi^+ p \to \pi^+\pi^+ n$, and $\pi^+ p \to \pi^+\pi^0 p$
in the energy region around $W=1.7$ GeV~\cite{j-parc}.
In what follows, we present predictions based on our 2016 model 
for the invariant mass distributions within the energy region
anticipated for this experiment.

\begin{figure}[t]
\begin{center}
\includegraphics[width=0.75\textwidth,angle=0,clip]{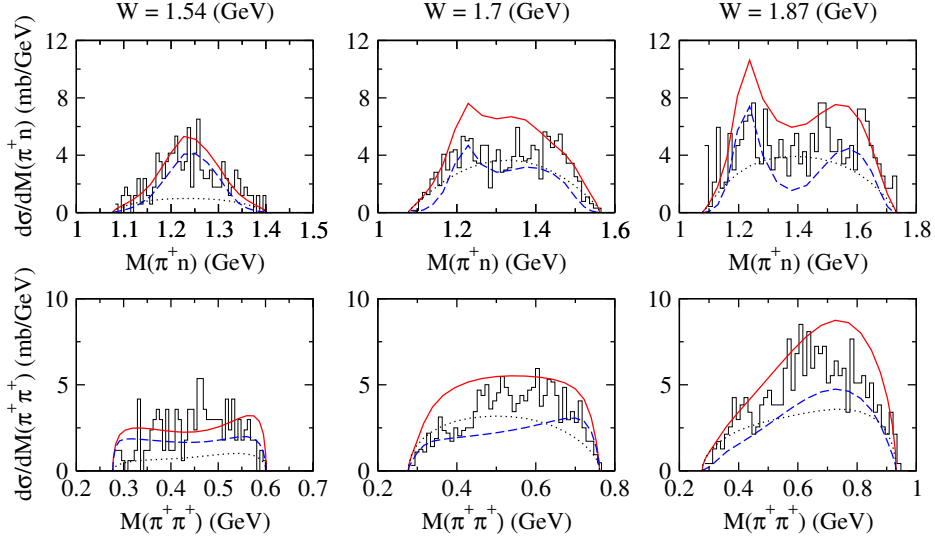}
\caption{
Invariant mass distributions of $\pi^+ p \to \pi^+ \pi^+ n$ at $W=1.54$, $1.70$, and~$1.87$~GeV.
The meaning of each curve is the same as in Fig.~\ref{fig:tcrs}.
The data are from Arndt~\cite{arndt}.
}
\label{fig:dsdm-pippipn}
\end{center}
\end{figure}
\begin{figure}[t]
\begin{center}
\includegraphics[width=0.75\textwidth,angle=0,clip]{dsdm-pippi0p.eps}
\caption{
Invariant mass distributions of $\pi^+ p \to \pi^+ \pi^0 p$ at $W=1.54$, $1.70$, and~$1.87$~GeV.
The meaning of each curve is the same as in Fig.~\ref{fig:tcrs}.
The data are from Arndt~\cite{arndt}.
}
\label{fig:dsdm-pippi0p}
\end{center}
\end{figure}
Figures~\ref{fig:dsdm-pippipn} and~\ref{fig:dsdm-pippi0p} show
the invariant mass distributions for 
$\pi^+ p \to \pi^+\pi^+ n$ and $\pi^+ p \to \pi^+\pi^0 p$
in the higher energy region at $W=1.54$, $1.7$, and $1.87$~GeV,
respectively.
Our predictions reproduce the overall behavior of 
the existing histogram data~\cite{arndt} well, although
they slightly (significantly) overestimate the data for
$\pi^+ p \to \pi^+\pi^+ n$ ($\pi^+ p \to \pi^+\pi^0 p$)
at $W=1.7$ and $1.87$~GeV.
Since the $\pi N \to \pi N$ subprocess in $T^{\textrm{dir}}$ 
[$T_{\pi N,\pi N}$ in the second term of Eq.~(\ref{eq:tdir2})] 
is already well determined within our 2016 model,
fine-tuning the $\Delta^* \to \pi\Delta$ vertex in $T^{\pi\Delta}$
will be required to reduce the predicted magnitudes and achieve 
a better description of the data.
This conclusion also holds for the $\pi^+ p \to \pi^+\pi^0 p$ reaction.

\begin{figure}[t]
\begin{center}
\includegraphics[width=0.75\textwidth,angle=0,clip]{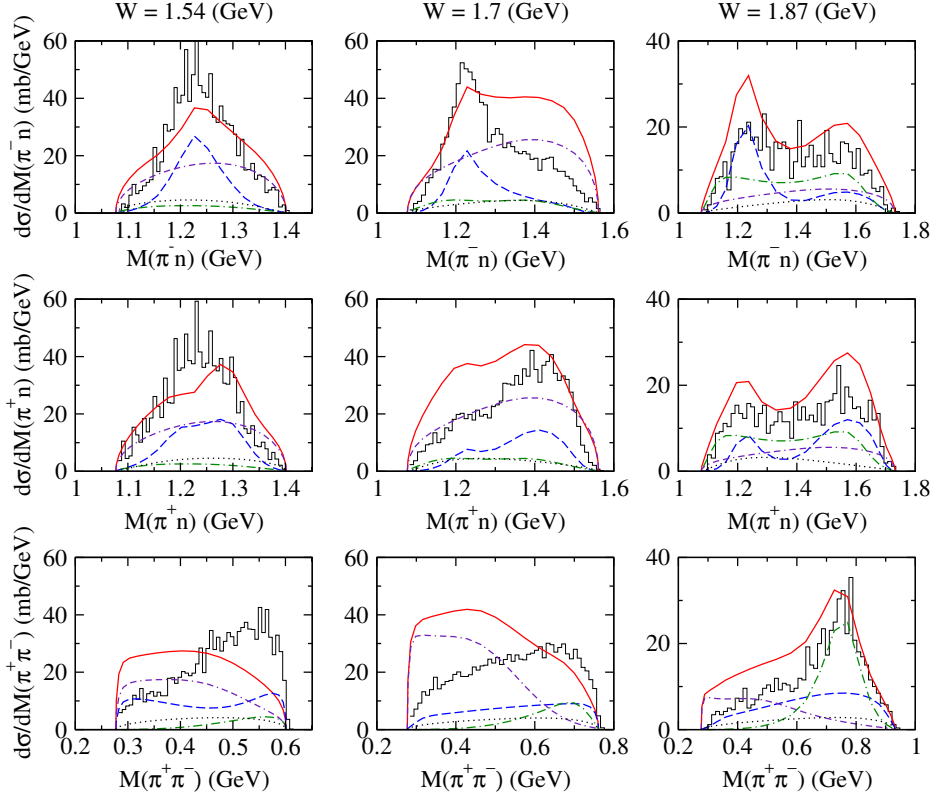}
\caption{
Invariant mass distributions of $\pi^- p \to \pi^+ \pi^- n$ at $W=1.54$, $1.70$, and~$1.87$~GeV.
The meaning of each curve is the same as in Fig.~\ref{fig:tcrs}
The data are from Arndt~\cite{arndt}.
}
\label{fig:dsdm-pippimn}
\end{center}
\end{figure}
\begin{figure}[t]
\begin{center}
\includegraphics[width=0.75\textwidth,angle=0,clip]{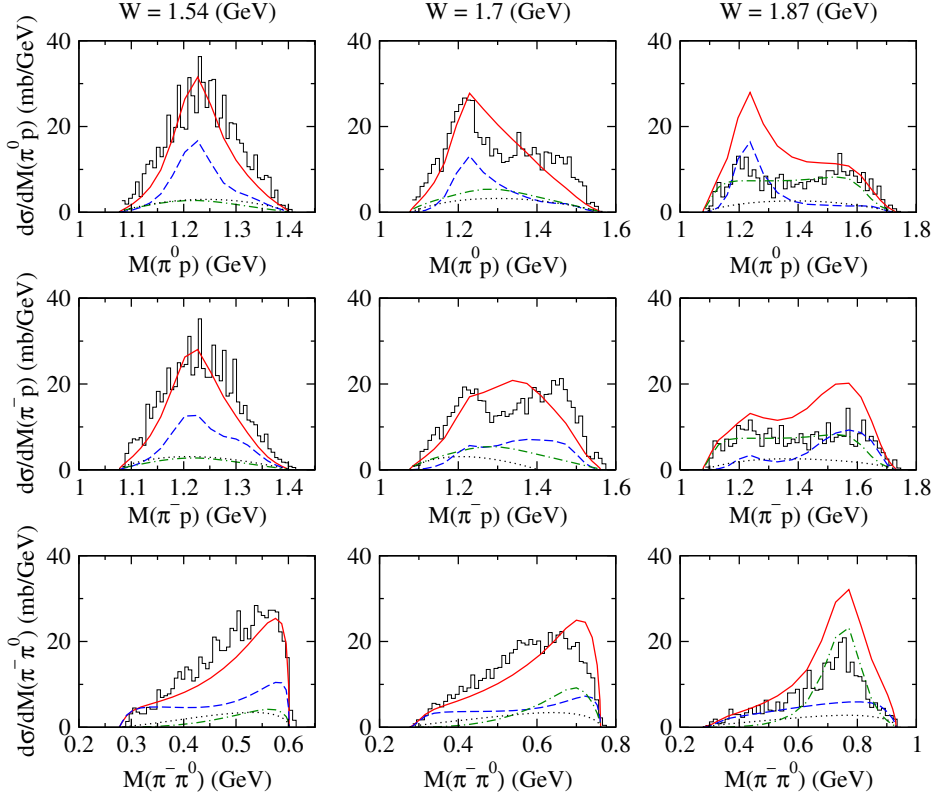}
\caption{
Invariant mass distributions of $\pi^- p \to \pi^- \pi^0 p$ at $W=1.54$, $1.70$, and~$1.87$~GeV.
The meaning of each curve is the same as in Fig.~\ref{fig:tcrs}
The data are from Arndt~\cite{arndt}.
}
\label{fig:dsdm-pimpi0p}
\end{center}
\end{figure}
Figures~\ref{fig:dsdm-pippimn} and~\ref{fig:dsdm-pimpi0p} show
the invariant mass distributions for 
$\pi^- p \to \pi^+\pi^- n$ and $\pi^- p \to \pi^-\pi^0 p$
at $W=1.54$, $1.7$, and $1.87$~GeV, respectively.
In contrast to the $\pi^+ p \to \pi \pi N$ case,
our results show discrepancies with the existing data 
not only in magnitude but also in shape.
For $\pi^- p \to \pi^+\pi^- n$,
the major source of these discrepancies 
appears to be the large contribution
from the $T^{\sigma N}$ process, implying that refinements to 
the $N^*\to \sigma N$ couplings are necessary.
On the other hand, the $T^{\sigma N}$ process does not contribute to 
the $\pi^- p \to \pi^-\pi^0 p$ reaction; 
thus, fine-tuning of all involved processes
($T^{\pi\Delta}$, $T^{\rho N}$, and $T^{\textrm{dir}}$),
including the interferences among them, will be required.

\begin{figure}[t]
\begin{center}
\includegraphics[width=0.75\textwidth,angle=0,clip]{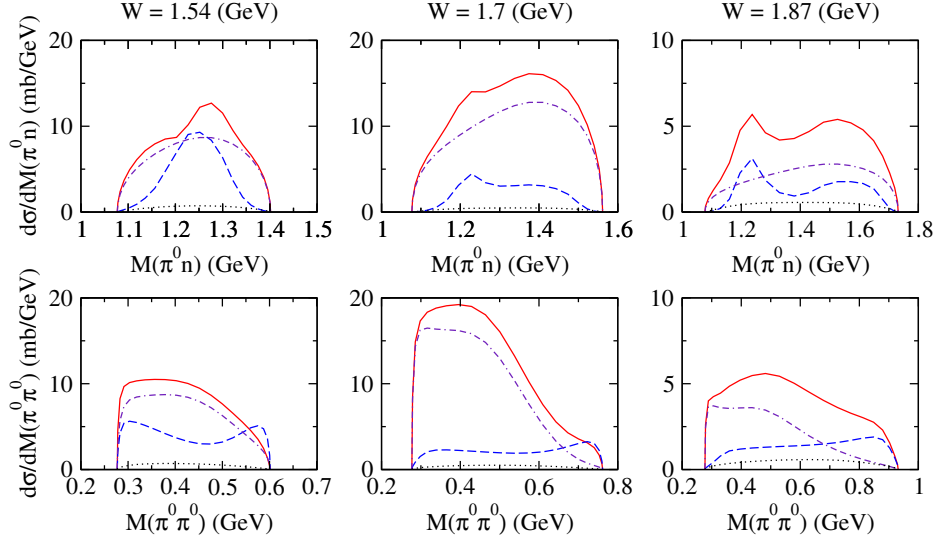}
\caption{
Invariant mass distributions of $\pi^- p \to \pi^0 \pi^0 n$ at $W=1.54$, $1.70$, and~$1.87$~GeV.
The meaning of each curve is the same as in Fig.~\ref{fig:tcrs}.
}
\label{fig:dsdm-pi0pi0n}
\end{center}
\end{figure}
Figure~\ref{fig:dsdm-pi0pi0n} shows
the invariant mass distributions for 
$\pi^- p \to \pi^0\pi^0 n$ at $W=1.54$, $1.7$, and $1.87$~GeV.
Although no data exist for this reaction at these energies,
we anticipate that a similar argument to that for the $\pi^- p \to \pi^+\pi^- n$ reaction
discussed above applies here as well.

\subsection{\label{sec:sensitivity}Sensitivity to the $N^*$ states}
We have seen from Figs.~\ref{fig:tcrs} 
and~\ref{fig:dsdm-pippipn}-\ref{fig:dsdm-pimpi0p} that,
in the energy region above $W\sim 1.65$~GeV,
which lies beyond the range covered by
the Crystal Ball and HADES experiments,
significant discrepancies between
our predictions and the existing data
appear for all five reactions.
This is mainly attributed to the fact that
the $\pi\Delta$, $\rho N$, and $\sigma N$ couplings to $N^*$
are not well determined by the data from single-meson production reactions
that were included in the partial-wave analysis to
construct our 2016 model.

In this subsection, we investigate
the sensitivity of the observables to variations of the
$\pi\Delta$, $\rho N$, and $\sigma N$ couplings to $N^*$.
Through this investigation, we aim to demonstrate which reaction
channels, observables, and kinematics
are useful for extracting $N^*$ parameters, thereby providing valuable
input for future experiments.
To proceed, we adopt the following procedure:
\begin{enumerate}
\item{
Recall that, within the ANL-Osaka DCC model,
the full partial-wave amplitude [Eq.~(\ref{eq:cc-eq})] for
$\pi N \to MB$ ($MB = \pi\Delta$, $\rho N$, $\sigma N$)
for a given total isospin $I$, total spin $J$, and parity $P$
can be expressed as a sum of
non-resonant ($t^{\textrm{NR}}$) and resonant ($t^{\textrm{R}}$) amplitudes
(here the energy and momentum indices are suppressed):
\begin{equation}
T_{MB(LS),\pi N} = 
 t^{\textrm{NR}}_{MB(LS),\pi N} 
+t^{\textrm{R}}_{MB(LS),\pi N},
\end{equation}
where $L$ and $S$ represent the allowed total angular momentum and spin
of the $MB$ channel for a given $(I, J, P)$, respectively.
The resonant amplitude $t^{\textrm{R}}$ is given by
\begin{equation}
t^{\textrm{R}}_{MB(LS),\pi N} = 
\sum_{n,m}
\bar{\Gamma}_{MB(LS),N^*_n}
D_{n,m}
\bar{\Gamma}_{N^*_m,\pi N},
\label{eq:resamp}
\end{equation}
where the indices $n$, $m$ specify the $n$-th and $m$-th bare $N^*$ states
with quantum numbers of $(I, J, P)$;
$\bar{\Gamma}_{MB(LS),N^*_n}$ ($\bar{\Gamma}_{N^*_m,\pi N}$)
is the fully dressed vertex for the $N^*_n \to MB(LS)$ ($\pi N \to N^*_m$)
transition; and 
$D_{n,m}$ is the dressed $N^*$ propagator~\cite{ao13-1}.
It is important to note that the dressed $N^*$ propagator
in the resonant amplitude $t^{\textrm{R}}$
contains the poles of \textit{physical} $N^*$ resonances obtained by solving
the full scattering equation [Eq.~(\ref{eq:cc-eq})],
and that the fully dressed vertices evaluated at these pole energies
can be associated with the ``coupling strength'' of the corresponding $N^\ast$ resonances
to the $MB$ states.\footnote{
Poles may be dynamically generated in $t^{\textrm{NR}}$
through the infinite iteration of the non-resonant potentials
[the first two terms on the right-hand side of Eq.~(\ref{eq:cc-v})].
However, if the bare $N^*$ states
[the last term on the right-hand side of Eq.~(\ref{eq:cc-v})]
are added to the transition potential $V_{\beta,\alpha}$,
the poles in $t^{\textrm{NR}}$ and the bare $N^*$ states
play a qualitatively similar role: the mixing between
the poles in $t^{\textrm{NR}}$ and the bare $N^*$ states
results in the physical resonances.
}
}
\item{
Multiply $\bar{\Gamma}_{MB(LS),N^*_n}$
by a common real factor $a_{MB(LS),N^*(I,J^P)}$ for all $n$
to define a modified fully dressed vertex:
\begin{equation}
\bar{\Gamma}'_{MB(LS),N^*_n} \equiv
a_{MB(LS),N^*(I,J^P)} \times \bar{\Gamma}_{MB(LS),N^*_n} \quad (\textrm{for all}~n),
\label{eq:modvtx}
\end{equation}
and replace $\bar{\Gamma}_{MB(LS),N^*_n}$ in the resonant amplitude $t^{\textrm{R}}$ [Eq.~(\ref{eq:resamp})]
with this modified one.
}
\item{
Compute the cross sections
by varying $a_{MB(LS),N^*(I,J^P)}$ from 1
and examine how the cross sections change as 
$a_{MB(LS),N^*(I,J^P)}$ is varied.
}
\end{enumerate}
The above procedure essentially corresponds to
uniformly varying the magnitudes of the $MB$ couplings 
to the physical $N^*$ resonances within a given partial wave.
This analysis serves to fulfill the objective described above.

In what follows, we investigate the sensitivity of 
the $\pi N \to \pi \pi N$ observables to the $N^*$ resonances
by examining the total cross sections [Eq.~(\ref{eq:tcrs})] and 
invariant mass distributions [Eq.~(\ref{eq:dsdm})].
The results for the angular distributions [Eq.~(\ref{eq:dsdx})] 
are presented in Appendix~\ref{sec:sensitivity2}.

\subsubsection{Isospin $I= 3/2$ $\Delta^*$}
We first investigate the sensitivity of the cross sections to the $I=3/2$ $\Delta^*$ resonances.
As we have seen from Figs.~\ref{fig:tcrs-pp-pwa} and~\ref{fig:tcrs-mp-pwa},
above $W=1.6$~GeV, $S_{31}$, $P_{33}$, $D_{33}$, and $F_{37}$ have large contributions
among the $I=3/2$ partial waves.

\begin{figure}[t]
\begin{center}
\includegraphics[width=\textwidth,angle=0,clip]{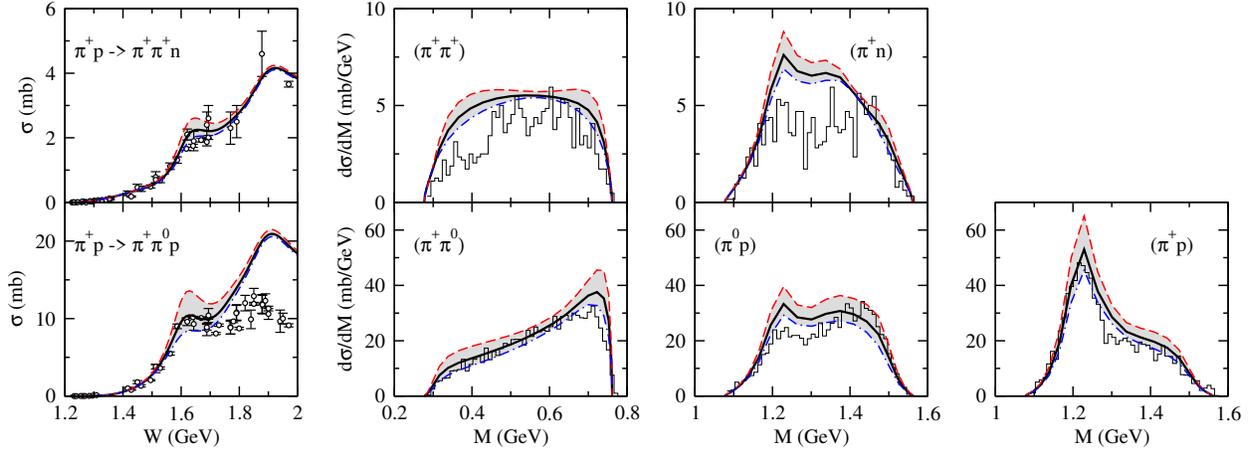}
\caption{
Sensitivity of the cross sections to 
the $S_{31}[\Delta^*(J^P=\frac{1}{2}^-)] \to \pi\Delta(L=2,S=\frac{3}{2})$ coupling.
Leftmost column: Total cross sections. 
Remaining columns: Invariant mass distributions for the corresponding reactions at $W=1.7$~GeV.
The solid curves represent the full results, while the grey bands show the range obtained 
by allowing a 50\% variation in the magnitude of the dressed vertices for
$S_{31}[\Delta^*(J^P=\frac{1}{2}^-)] \to \pi\Delta(L=2,S=\frac{3}{2})$ 
(the red dashed and blue dash-dot curves correspond to the results with 
$a_{MB(LS),N^*(I,J^P)} = 1.5$ and~$0.5$, respectively).
The data are taken from Refs.~\cite{manley,ao09-1,arndt} and references therein.
}
\label{fig:S31-piD1-1700}
\end{center}
\end{figure}
\begin{figure}[t]
\begin{center}
\includegraphics[width=\textwidth,angle=0,clip]{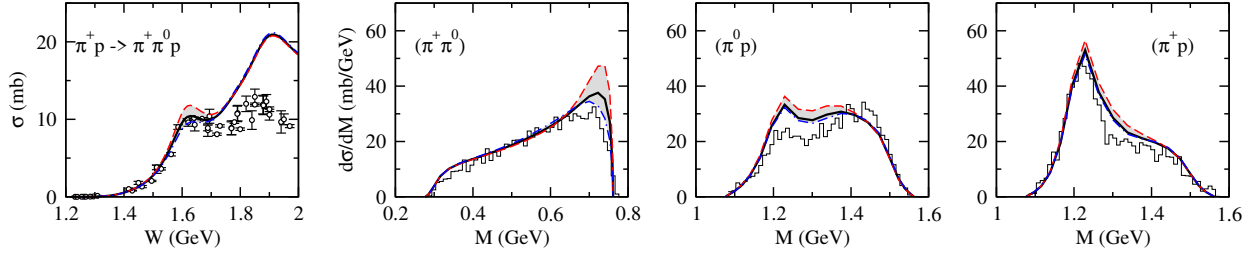}
\caption{
Sensitivity of the cross sections to 
the $S_{31}[\Delta^*(J^P=\frac{1}{2}^-)] \to \rho N(L=0,S=\frac{1}{2})$ coupling.
Leftmost column: Total cross sections. 
Remaining columns: Invariant mass distributions for the corresponding reactions at $W=1.7$~GeV.
The solid curves represent the full results, while the grey bands show the range obtained 
by allowing a 50\% variation in the magnitude of the dressed vertices for
$S_{31}[\Delta^*(J^P=\frac{1}{2}^-)] \to \rho N(L=0,S=\frac{1}{2})$ 
(the red dashed and blue dash-dot curves correspond to the results with 
$a_{MB(LS),N^*(I,J^P)} = 1.5$ and~$0.5$, respectively).
The data are taken from Refs.~\cite{manley,ao09-1,arndt} and references therein.
}
\label{fig:S31-rhoN1-1700}
\end{center}
\end{figure}
$\bm{S_{31}}$:
Figures~\ref{fig:S31-piD1-1700} and~\ref{fig:S31-rhoN1-1700}
show the sensitivity of the total cross sections and invariant mass distributions to 
the strengths of the dressed vertices for 
$S_{31}[\Delta^*(J^P=\frac{1}{2}^-)] \to \pi\Delta(L=2,S=\frac{3}{2})$ and
$S_{31}[\Delta^*(J^P=\frac{1}{2}^-)] \to \rho N(L=0,S=\frac{1}{2})$, respectively.
Here, we present results only for the reactions that are sensitive to 
the $\pm 50\%$ variation in the magnitudes of these vertices.
Results for other dressed vertices are also omitted due to their negligible sensitivity.
(We apply this same criterion to the discussions of other partial waves below.)
In the total cross sections, we observe that these variations
have a significant effect at $W=1.6$--$1.7$~GeV.
This corresponds to the energy region where
two resonances with pole masses of $1597-i69$~MeV and $1713-i187$~MeV
are located in this partial wave within our 2016 model~\cite{ao16}.
Regarding the invariant mass distributions at $W=1.7$~GeV, the variation in
the $S_{31}[\Delta^*(J^P=\frac{1}{2}^-)] \to \pi\Delta(L=2,S=\frac{3}{2})$ vertex
affects the $\pi^+ p \to \pi^+ \pi^+ n$ and $\pi^+ p \to \pi^+ \pi^0 p$ reactions
uniformly over the entire kinematically allowed range.
In contrast, the impact of the $S_{31}[\Delta^*(J^P=\frac{1}{2}^-)] \to \rho N(L=0,S=\frac{1}{2})$ variation
is generally small, although a significant effect is observed in the $\pi^+\pi^0$ invariant mass distribution
for the $\pi^+ p \to \pi^+ \pi^0 p$ reaction in the region of $M_{\pi\pi} \gtrsim 0.65$~GeV.
A reduction in the magnitude of the $S_{31}[\Delta^*(J^P=\frac{1}{2}^-)] \to \rho N(L=0,S=\frac{1}{2})$ vertex
appears to be favored by the data.

\begin{figure}[t]
\begin{center}
\includegraphics[width=\textwidth,angle=0,clip]{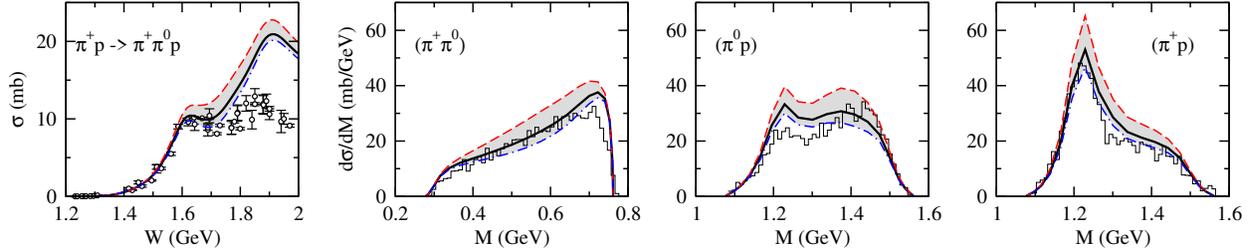}
\caption{
Sensitivity of the cross sections to 
the $P_{33}[\Delta^*(J^P=\frac{3}{2}^+)] \to \pi\Delta(L=1,S=\frac{3}{2})$ coupling.
Leftmost column: Total cross sections. 
Remaining columns: Invariant mass distributions for the corresponding reactions at $W=1.7$~GeV.
The solid curves represent the full results, while the grey bands show the range obtained 
by allowing a 50\% variation in the magnitude of the dressed vertices for
$P_{33}[\Delta^*(J^P=\frac{3}{2}^+)] \to \pi\Delta(L=1,S=\frac{3}{2})$ 
(the red dashed and blue dash-dot curves correspond to the results with 
$a_{MB(LS),N^*(I,J^P)} = 1.5$ and~$0.5$, respectively).
The data are taken from Refs.~\cite{manley,ao09-1,arndt} and references therein.
}
\label{fig:P33-piD1-1700}
\end{center}
\end{figure}
$\bm{P_{33}}$:
Figure~\ref{fig:P33-piD1-1700}
shows the sensitivity of the total cross sections and invariant mass distributions to 
the strength of the dressed vertex for 
$P_{33}[\Delta^*(J^P=\frac{3}{2}^+)] \to \pi\Delta(L=1,S=\frac{3}{2})$.
A visible effect from the $\pm 50\%$ variation in
the $P_{33}[\Delta^*(J^P=\frac{3}{2}^+)] \to \pi\Delta(L=1,S=\frac{3}{2})$ vertex
is observed in the total cross sections over a wide energy region above $W \sim 1.6$~GeV.
However, this effect appears minor and is not sufficient to improve the description of the data.
This behavior would be attributed to the broad $P_{33}$ resonance with a pole mass of
$1733-i162$~MeV found in our 2016 model~\cite{ao16}.
The influence on the invariant mass distributions at $W=1.7$~GeV is also visible but minor.

\begin{figure}[t]
\begin{center}
\includegraphics[width=\textwidth,angle=0,clip]{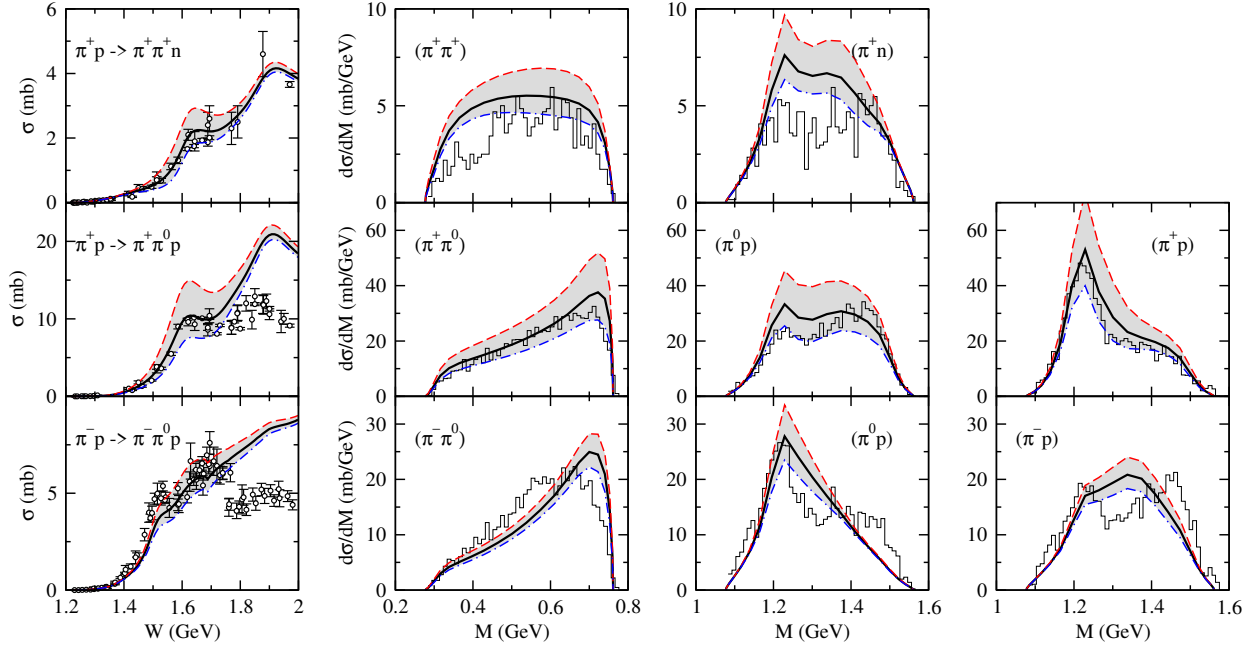}
\caption{
Sensitivity of the cross sections to 
the $D_{33}[\Delta^*(J^P=\frac{3}{2}^-)] \to \pi\Delta(L=0,S=\frac{3}{2})$ coupling.
Leftmost column: Total cross sections. 
Remaining columns: Invariant mass distributions for the corresponding reactions at $W=1.7$~GeV.
The solid curves represent the full results, while the grey bands show the range obtained 
by allowing a 50\% variation in the magnitude of the dressed vertices for
$D_{33}[\Delta^*(J^P=\frac{3}{2}^-)] \to \pi\Delta(L=0,S=\frac{3}{2})$ 
(the red dashed and blue dash-dot curves correspond to the results with 
$a_{MB(LS),N^*(I,J^P)} = 1.5$ and~$0.5$, respectively).
The data are taken from Refs.~\cite{manley,ao09-1,arndt} and references therein.
}
\label{fig:D33-piD1-1700}
\end{center}
\end{figure}
$\bm{D_{33}}$:
Figure~\ref{fig:D33-piD1-1700}
shows the sensitivity of the total cross sections and invariant mass distributions to 
the strength of the dressed vertex for 
$D_{33}[\Delta^*(J^P=\frac{3}{2}^-)] \to \pi\Delta(L=0,S=\frac{3}{2})$.
The effect of the $\pm 50\%$ variation in
the magnitude of the $D_{33}[\Delta^*(J^P=\frac{3}{2}^-)] \to \pi\Delta(L=0,S=\frac{3}{2})$ vertex 
becomes visible above $W \sim 1.5$~GeV and significantly large around $W=1.6$--$1.7$~GeV.
This is consistent with the existence of a resonance with a pole mass of $1577-i113$~MeV
in this partial wave within our 2016 model~\cite{ao16}.
In contrast to the $S_{31}$ and $P_{33}$ cases, variations in this partial wave 
show a visible influence on $\pi^- p \to \pi^- \pi^0 p$ as well.
The variation in the magnitude of 
the $D_{33}[\Delta^*(J^P=\frac{3}{2}^-)] \to \pi\Delta(L=0,S=\frac{3}{2})$ vertex 
affects the invariant mass distributions at $W=1.7$~GeV
uniformly over the entire kinematically allowed range, 
but does not significantly change the shape of the distributions.

\begin{figure}[t]
\begin{center}
\includegraphics[width=\textwidth,angle=0,clip]{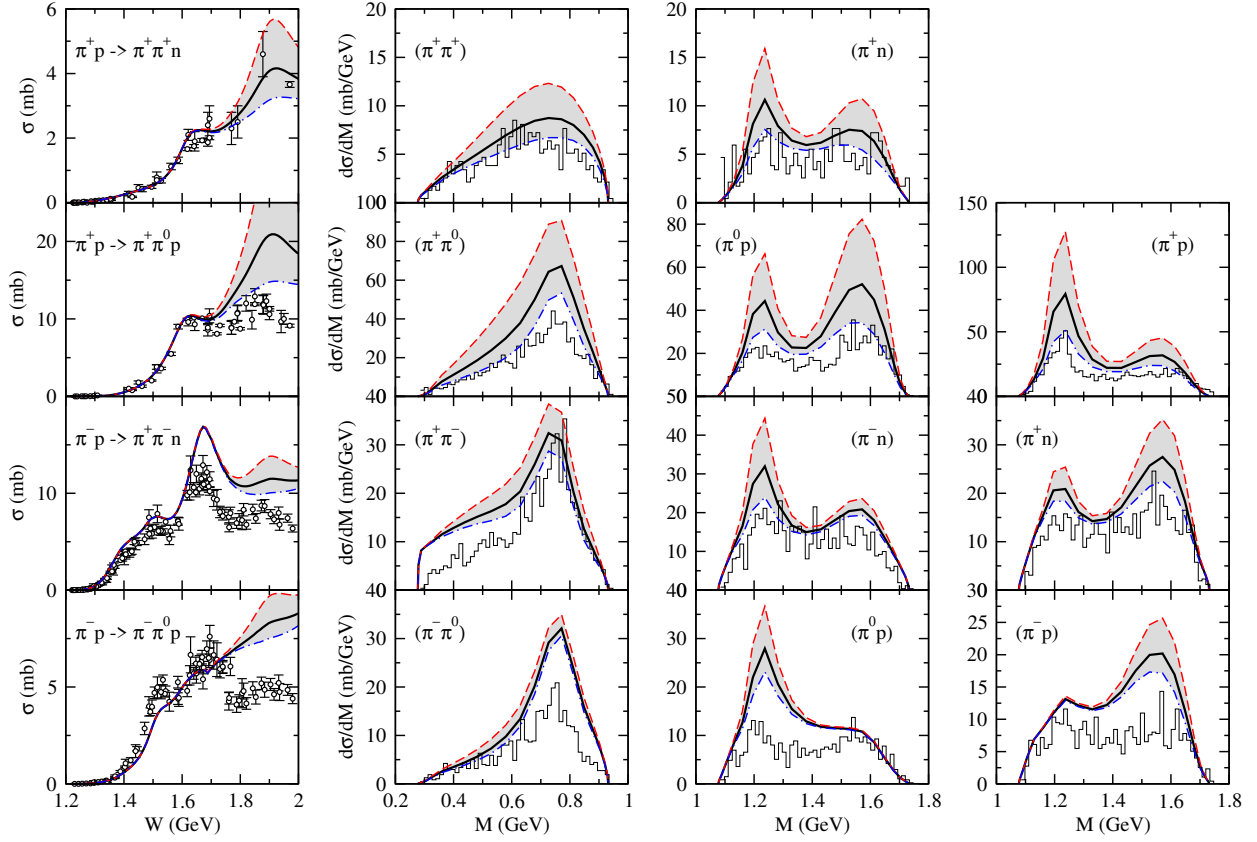}
\caption{
Sensitivity of the cross sections to 
the $F_{37}[\Delta^*(J^P=\frac{7}{2}^+)] \to \pi\Delta(L=3,S=\frac{3}{2})$ coupling.
Leftmost column: Total cross sections. 
Remaining columns: Invariant mass distributions for the corresponding reactions at $W=1.87$~GeV.
The solid curves represent the full results, while the grey bands show the range obtained 
by allowing a 50\% variation in the magnitude of the dressed vertices for
$F_{37}[\Delta^*(J^P=\frac{7}{2}^+)] \to \pi\Delta(L=3,S=\frac{3}{2})$ 
(the red dashed and blue dash-dot curves correspond to the results with 
$a_{MB(LS),N^*(I,J^P)} = 1.5$ and~$0.5$, respectively).
The data are taken from Refs.~\cite{manley,ao09-1,arndt} and references therein.
}
\label{fig:F37-piD1-1870}
\end{center}
\end{figure}
$\bm{F_{37}}$:
Figure~\ref{fig:F37-piD1-1870} shows
the sensitivity of the total cross sections and invariant mass distributions to 
the strength of the dressed vertex for 
$F_{37}[\Delta^*(J^P=\frac{7}{2}^+)] \to \pi\Delta(L=3,S=\frac{3}{2})$.
In our 2016 model~\cite{ao16}, a single resonance with a pole mass of $1885-i102$~MeV 
was found for the $F_{37}$ partial wave, and indeed this partial wave shows 
a significant contribution to the cross sections
around the energy region of this resonance pole 
(see Figs.~\ref{fig:tcrs-pp-pwa} and~\ref{fig:tcrs-mp-pwa}).
Varying the magnitude of the $F_{37}[\Delta^*(J^P=\frac{7}{2}^+)] \to \pi\Delta(L=3,S=\frac{3}{2})$
dressed vertex by $\pm 50\%$ significantly affects the cross sections except $\pi^- p \to \pi^0 \pi^0 n$,
and a reduction in this vertex strength is favored by the data.
Conversely, we find that the cross sections are insensitive to variations 
in the other decay channels, 
$F_{37}[\Delta^*(J^P=\frac{7}{2}^+)] \to \pi\Delta(L=5,S=\frac{3}{2})$ and 
$F_{37}[\Delta^*(J^P=\frac{7}{2}^+)] \to \rho N$.
This observation seems consistent with the findings in Sec.~IV of Ref.~\cite{ao13-2}.
While reducing the magnitude of the $F_{37}[\Delta^*(J^P=\frac{7}{2}^+)] \to \pi\Delta(L=3,S=\frac{3}{2})$ vertex
is favorable for reproducing the $\pi N \to \pi\pi N$ data,
the resulting impact on the cross sections of other reactions, 
such as $\pi N \to MB$ ($MB=\pi N,K\Sigma$), through coupled-channel effects,
can be compensated by adjusting the $F_{37}[\Delta^*(J^P=\frac{7}{2}^+)] \to \rho N$ couplings,
which have a negligible influence on the $\pi N \to \pi \pi N$ reaction.

\subsubsection{Isospin $I= 1/2$ $N^*$}
Next we investigate the sensitivity of the cross sections to the $I=1/2$ $N^*$ resonances.
As we have seen from Fig.~\ref{fig:tcrs-mp-pwa},
above $W=1.6$~GeV, $D_{13}$, $D_{15}$, and $F_{15}$ have large contributions
among the $I=1/2$ partial waves.

\begin{figure}[t]
\begin{center}
\includegraphics[width=\textwidth,angle=0,clip]{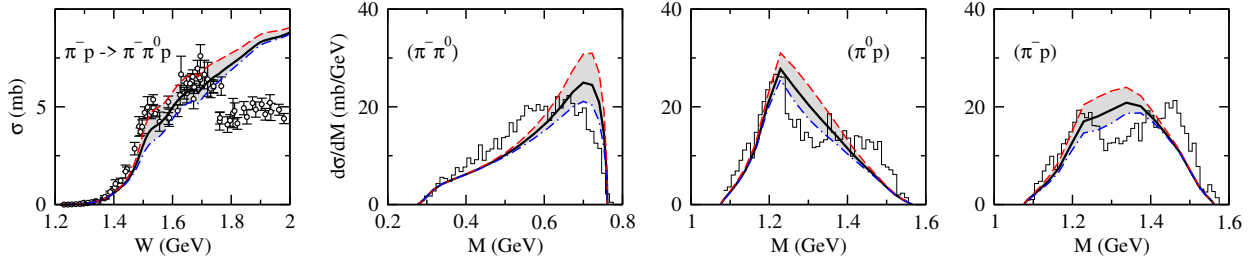}
\caption{
Sensitivity of the cross sections to 
the $D_{13}[N^*(J^P=\frac{3}{2}^-)] \to \rho N(L=0,S=\frac{3}{2})$ coupling.
Leftmost column: Total cross sections. 
Remaining columns: Invariant mass distributions for the corresponding reactions at $W=1.7$~GeV.
The solid curves represent the full results, while the grey bands show the range obtained 
by allowing a 50\% variation in the magnitude of the dressed vertices for
$D_{13}[N^*(J^P=\frac{3}{2}^-)] \to \rho N(L=0,S=\frac{3}{2})$ 
(the red dashed and blue dash-dot curves correspond to the results with 
$a_{MB(LS),N^*(I,J^P)} = 1.5$ and~$0.5$, respectively).
The data are taken from Refs.~\cite{manley,ao09-1,arndt} and references therein.
}
\label{fig:D13-rhoN2-1700}
\end{center}
\end{figure}
$\bm{D_{13}}$:
Figure~\ref{fig:D13-rhoN2-1700} shows
the sensitivity of the total cross sections and invariant mass distributions to 
the strength of the dressed vertex for 
$D_{13}[N^*(J^P=\frac{3}{2}^-)] \to \rho N(L=0,S=\frac{3}{2})$.
We find that only the $\pi^- p \to \pi^- \pi^0 p$ reaction is sensitive to the variation in
the magnitude of $D_{13}[N^*(J^P=\frac{3}{2}^-)] \to \rho N(L=0,S=\frac{3}{2})$ vertex.
The total cross sections at $W=1.5$--$1.7$~GeV are sensitive to the variation of this vertex,
which seems consistent with the existence of two
resonances with pole masses of $1509-i48$~MeV and $1702-i148$~MeV\footnote{The second pole
is located in the complex energy plane slightly off the closest sheet to the physical real energy axis.}
found in our 2016 model~\cite{ao16} for this partial wave.
Similar to the $D_{33}[\Delta^*(J^P=\frac{3}{2}^-)] \to \pi\Delta(L=0,S=\frac{3}{2})$ case,
the variation in the magnitude of $D_{13}[N^*(J^P=\frac{3}{2}^-)] \to \rho N(L=0,S=\frac{3}{2})$ vertex
does not improve the shape of invariant mass distributions at $W=1.7$~GeV.

\begin{figure}[t]
\begin{center}
\includegraphics[width=\textwidth,angle=0,clip]{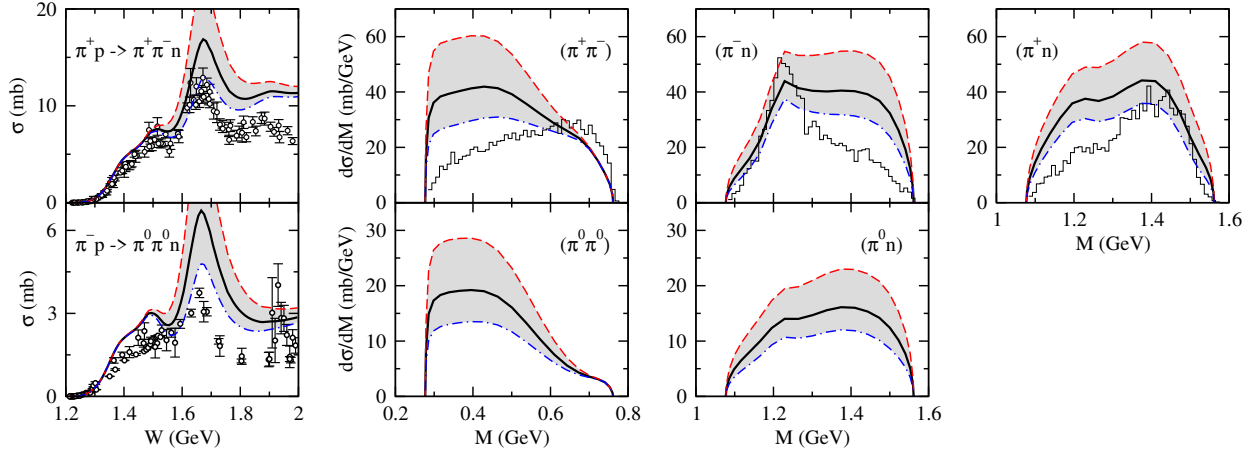}
\caption{
Sensitivity of the cross sections to 
the $D_{15}[N^*(J^P=\frac{5}{2}^-)] \to \sigma N(L=3,S=\frac{1}{2})$ coupling.
Leftmost column: Total cross sections. 
Remaining columns: Invariant mass distributions for the corresponding reactions at $W=1.7$~GeV.
The solid curves represent the full results, while the grey bands show the range obtained 
by allowing a 50\% variation in the magnitude of the dressed vertices for
$D_{15}[N^*(J^P=\frac{5}{2}^-)] \to \sigma N(L=3,S=\frac{1}{2})$ 
(the red dashed and blue dash-dot curves correspond to the results with 
$a_{MB(LS),N^*(I,J^P)} = 1.5$ and~$0.5$, respectively).
The data are taken from Refs.~\cite{manley,ao09-1,arndt} and references therein.
}
\label{fig:D15-sigN1-1700}
\end{center}
\end{figure}
$\bm{D_{15}}$:
Figure~\ref{fig:D15-sigN1-1700} shows
the sensitivity of the total cross sections and invariant mass distributions to 
the strength of the dressed vertex for 
$D_{15}[N^*(J^P=\frac{5}{2}^-)] \to \sigma N(L=3,S=\frac{1}{2})$.
We note that the $\pi N \to N^* \to \sigma N$ process contributes only 
to the $\pi^- p \to \pi^+ \pi^- n$ and $\pi^- p \to \pi^0 \pi^0 n$ reactions.
The results for the total cross sections and the invariant mass distributions at $W=1.7$~GeV 
show that the $50\%$ variation of the magnitude of this vertex has a significant effect
around $W=1.7$~GeV, which seems consistent with the existence of 
a resonance with a pole mass of $1651-i68$~MeV found in our 2016 model~\cite{ao16} for this partial wave.
Reducing the magnitude of the $D_{15}[N^*(J^P=\frac{5}{2}^-)] \to \sigma N(L=3,S=\frac{1}{2})$ vertex
seems to be favored by the data and will be key to resolving the overestimation of the cross sections around $W=1.7$~GeV.

\begin{figure}[t]
\begin{center}
\includegraphics[width=\textwidth,angle=0,clip]{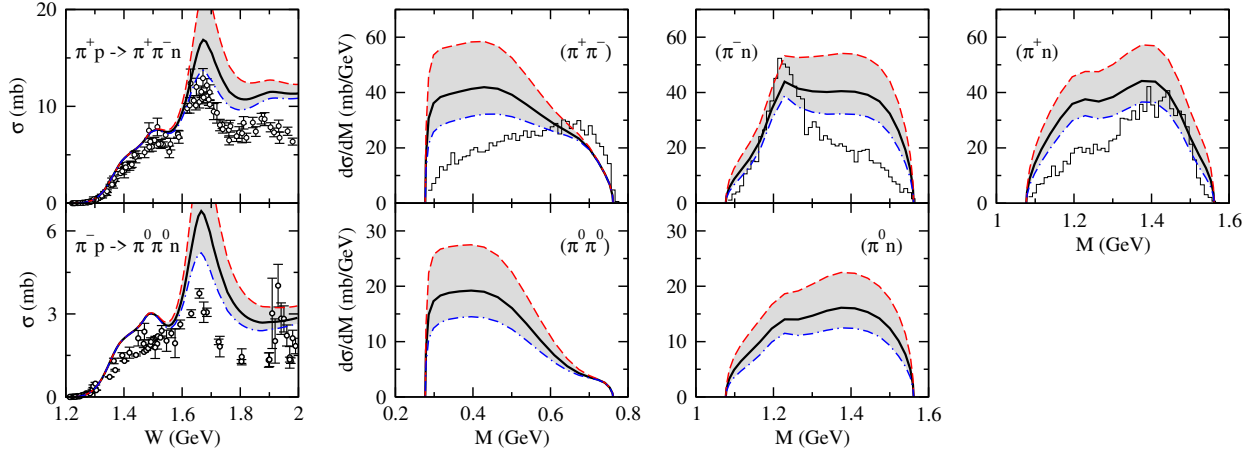}
\caption{
Sensitivity of the cross sections to 
the $F_{15}[N^*(J^P=\frac{5}{2}^+)] \to \sigma N(L=2,S=\frac{1}{2})$ coupling.
Leftmost column: Total cross sections. 
Remaining columns: Invariant mass distributions for the corresponding reactions at $W=1.7$~GeV.
The solid curves represent the full results, while the grey bands show the range obtained 
by allowing a 50\% variation in the magnitude of the dressed vertices for
$F_{15}[N^*(J^P=\frac{5}{2}^+)] \to \sigma N(L=2,S=\frac{1}{2})$ 
(the red dashed and blue dash-dot curves correspond to the results with 
$a_{MB(LS),N^*(I,J^P)} = 1.5$ and~$0.5$, respectively).
The data are taken from Refs.~\cite{manley,ao09-1,arndt} and references therein.
}
\label{fig:F15-sigN1-1700}
\end{center}
\end{figure}
$\bm{F_{15}}$:
Figure~\ref{fig:F15-sigN1-1700} shows
the sensitivity of the total cross sections and invariant mass distributions to 
the strength of the dressed vertex for 
$F_{15}[N^*(J^P=\frac{5}{2}^+)] \to \sigma N(L=2,S=\frac{1}{2})$.
This partial wave has a resonance with a pole mass of $1665-i52$~MeV found in our 2016 model~\cite{ao16},
and the $50\%$ variation in the magnitude of this vertex shows a behavior quite similar to
that of the $D_{15}[N^*(J^P=\frac{5}{2}^-)] \to \sigma N(L=3,S=\frac{1}{2})$ case.
Since the coupled-channel effects of the $\sigma N$ channel on the other $\pi N$ reactions
[$\pi N \to MB$ ($MB=\pi N, \eta N, K\Lambda, K\Sigma$)] are in general quite small,
simply reducing the magnitude of both the $F_{15}[N^*(J^P=\frac{5}{2}^+)] \to \sigma N(L=2,S=\frac{1}{2})$ 
and $D_{15}[N^*(J^P=\frac{5}{2}^-)] \to \sigma N(L=3,S=\frac{1}{2})$ vertices
is very promising to resolve the overestimation of the cross sections for 
$\pi^- p \to \pi^+ \pi^- n$ around $W=1.7$~GeV and
for $\pi^- p \to \pi^0 \pi^0 n$ above $W=1.6$~GeV.

\section{\label{sec:summary}Summary}
We have extended our previous investigations of the
$\pi N \to \pi \pi N$ reactions~\cite{ao09-1,ao13-2} 
to predict the cross sections for these processes
using the 2016 ANL-Osaka DCC model~\cite{ao16}.
The  model was constructed by fitting only the single-meson production
data for $\pi N$ and $\gamma N$ reactions:
$\pi N,\gamma N \to \pi N, \eta N, K\Lambda, K\Sigma$.
Thus, the results for $\pi N \to \pi \pi N$ presented here
are pure predictions of the model,
which serve to examine the extent to which the forthcoming data
from J-PARC can be described. 
This study provides information for improving
the extraction of nucleon resonances that have large decay
widths to $\pi \pi N$ states 
via the quasi-two-body $\pi\Delta$, $\rho N$, and $\sigma N$ channels.

We have presented the predicted total cross sections, 
invariant mass distributions, and angular distributions.
Our results reasonably reproduce the overall features of the existing data
even without including the $\pi N \to \pi \pi N$ reaction data in the fit,
although discrepancies are observed, 
most notably a significant overestimation of the cross sections above $W \sim 1.65$~GeV.
By examining the contributions from the 
$T^{\pi\Delta}$, $T^{\rho N}$, $T^{\sigma N}$, and $T^{\textrm{dir}}$ processes 
and individual partial waves,
we have identified the primary sources of these discrepancies.

We have also investigated the sensitivity of the cross sections
to the $\pi\Delta$, $\rho N$, and $\sigma N$ couplings of nucleon resonances
by varying the magnitudes of the fully dressed vertices for $N^* \to \pi\Delta,\rho N,\sigma N$
according to the prescription defined in Eq.~(\ref{eq:modvtx}).
Through this investigation, we have identified the observables and energy regions
where the high-mass resonances in the
$S_{31}$, $P_{33}$, $D_{33}$, $F_{37}$, $D_{13}$, $D_{15}$, and $F_{15}$ partial waves
can be most effectively investigated.

We hope that the comprehensive predictions and sensitivity analyses 
presented in this paper will serve as a useful guide for future experiments.
These insights are particularly relevant for the planning and analysis of 
the forthcoming measurements at J-PARC.

\acknowledgments
The authors would like to thank Kyungseon Joo for useful discussions, 
and Shin Hyung Kim and Sangho Kim for providing information on the experiment at J-PARC.
T.-S.H.L. is supported by the Office of Science of the U.S. Department of Energy under Contract No. DE-AC02-05CH112.

\appendix

\section{\label{sec:sensitivity2}Sensitivity of angular distributions to the $N^*$ states}
\begin{figure}[ht]
\begin{center}
\includegraphics[width=0.75\textwidth,angle=0,clip]{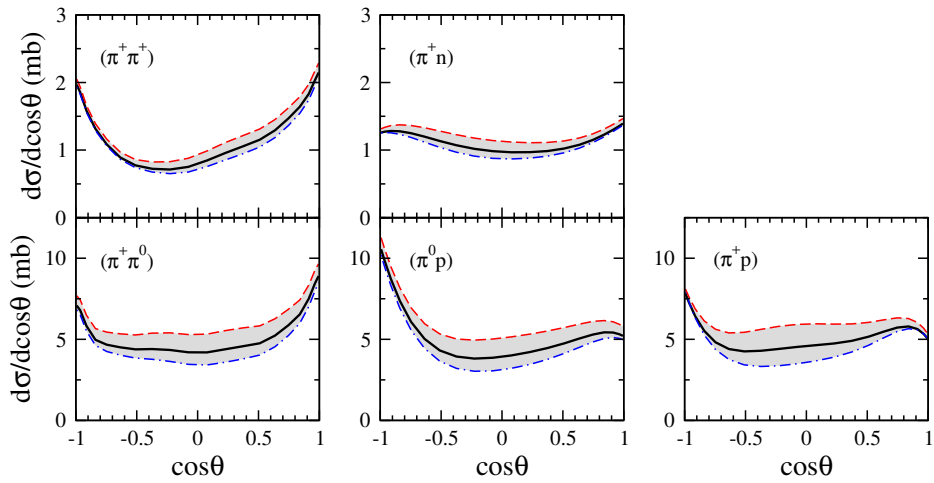}
\caption{
Sensitivity of the angular distributions at $W=1.7$~GeV to 
the $S_{31}[\Delta^*(J^P=\frac{1}{2}^-)] \to \pi\Delta(L=2,S=\frac{3}{2})$ coupling.
First row: $\pi^+ p \to \pi^+ \pi^+ n$;
second row: $\pi^+ p \to \pi^+ \pi^0 p$.
The scattering angle $\theta$ denotes the direction of the outgoing two-body subsystem 
indicated in parentheses.
The solid curves denote the full results. 
The grey bands indicate the uncertainty 
resulting from a 50\% variation in the $S_{31} \to \pi \Delta$ vertex magnitude 
(red dashed: $\times 1.5$; blue dash-dotted: $\times 0.5$).
}
\label{fig:S31-piD1-1700-dsdx}
\end{center}
\end{figure}

\begin{figure}[ht]
\begin{center}
\includegraphics[width=0.75\textwidth,angle=0,clip]{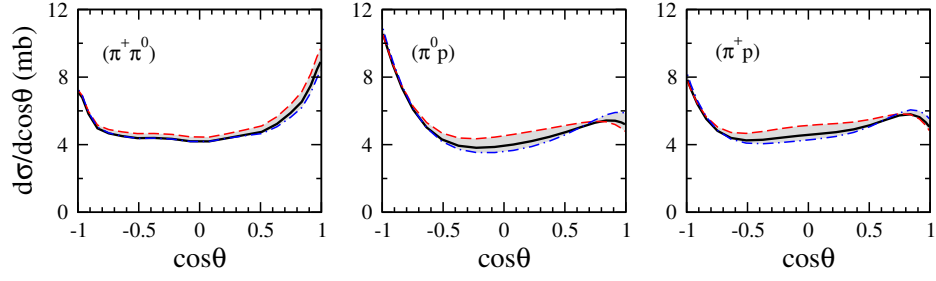}
\caption{
Sensitivity of the angular distributions for $\pi^+ p \to \pi^+ \pi^0 p$ at $W=1.7$~GeV to
the $S_{31}[\Delta^*(J^P=\frac{1}{2}^-)] \to \rho N(L=0,S=\frac{1}{2})$ coupling.
The scattering angle $\theta$ denotes the direction of the outgoing two-body subsystem 
indicated in parentheses.
The solid curves denote the full results. 
The grey bands indicate the uncertainty 
resulting from a 50\% variation in the $S_{31} \to \rho N$ vertex magnitude 
(red dashed: $\times 1.5$; blue dash-dotted: $\times 0.5$).
}
\label{fig:S31-rhoN1-1700-dsdx}
\end{center}
\end{figure}

\begin{figure}[ht]
\begin{center}
\includegraphics[width=0.75\textwidth,angle=0,clip]{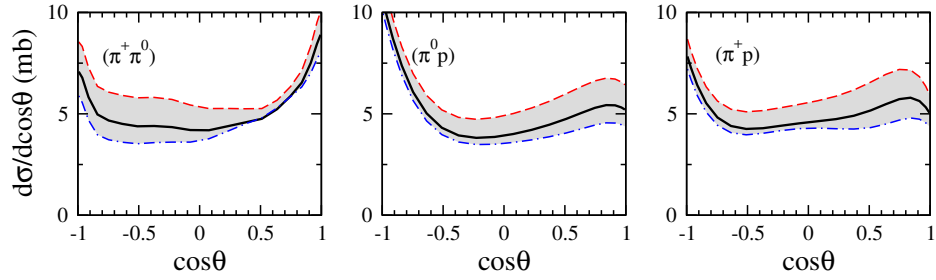}
\caption{
Sensitivity of the angular distributions for $\pi^+ p \to \pi^+ \pi^0 p$ at $W=1.7$~GeV to 
the $P_{33}[\Delta^*(J^P=\frac{3}{2}^+)] \to \pi\Delta(L=1,S=\frac{3}{2})$ coupling.
The scattering angle $\theta$ denotes the direction of the outgoing two-body subsystem 
indicated in parentheses.
The solid curves denote the full results. 
The grey bands indicate the uncertainty 
resulting from a 50\% variation in the $P_{33} \to \pi \Delta$ vertex magnitude 
(red dashed: $\times 1.5$; blue dash-dotted: $\times 0.5$).
}
\label{fig:P33-piD1-1700-dsdx}
\end{center}
\end{figure}

\begin{figure}[ht]
\begin{center}
\includegraphics[width=0.75\textwidth,angle=0,clip]{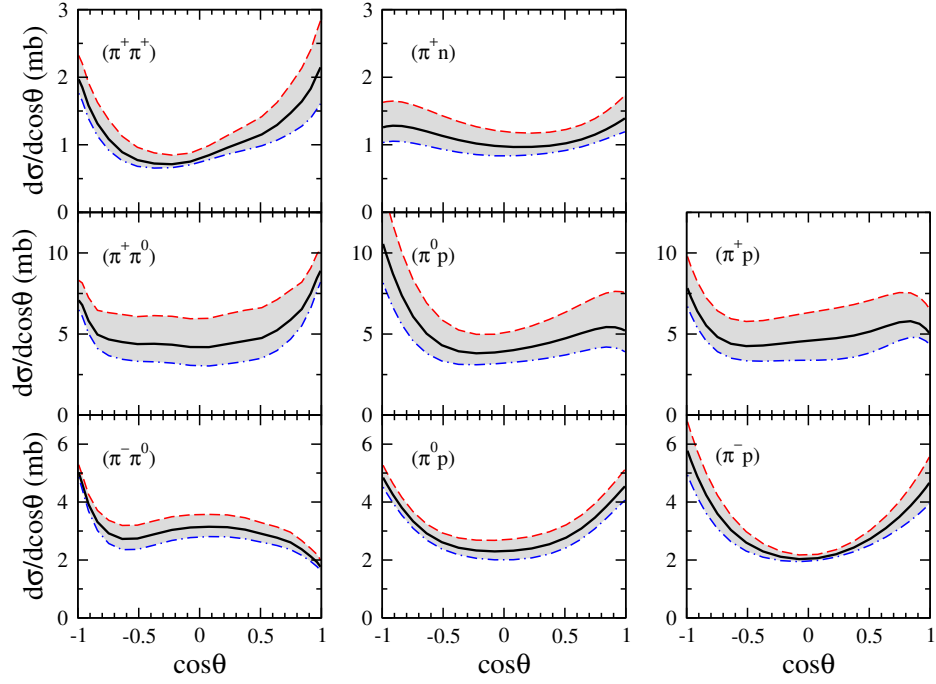}
\caption{
Sensitivity of the angular distributions at $W=1.7$~GeV to 
the $D_{33}[\Delta^*(J^P=\frac{3}{2}^-)] \to \pi\Delta(L=0,S=\frac{3}{2})$ coupling.
First row: $\pi^+ p \to \pi^+ \pi^+ n$;
second row: $\pi^+ p \to \pi^+ \pi^0 p$;
third row: $\pi^- p \to \pi^- \pi^0 p$.
The scattering angle $\theta$ denotes the direction of the outgoing two-body subsystem 
indicated in parentheses.
The solid curves denote the full results. 
The grey bands indicate the uncertainty 
resulting from a 50\% variation in the $D_{33} \to \pi \Delta$ vertex magnitude 
(red dashed: $\times 1.5$; blue dash-dotted: $\times 0.5$).
}
\label{fig:D33-piD1-1700-dsdx}
\end{center}
\end{figure}

\begin{figure}[ht]
\begin{center}
\includegraphics[width=0.75\textwidth,angle=0,clip]{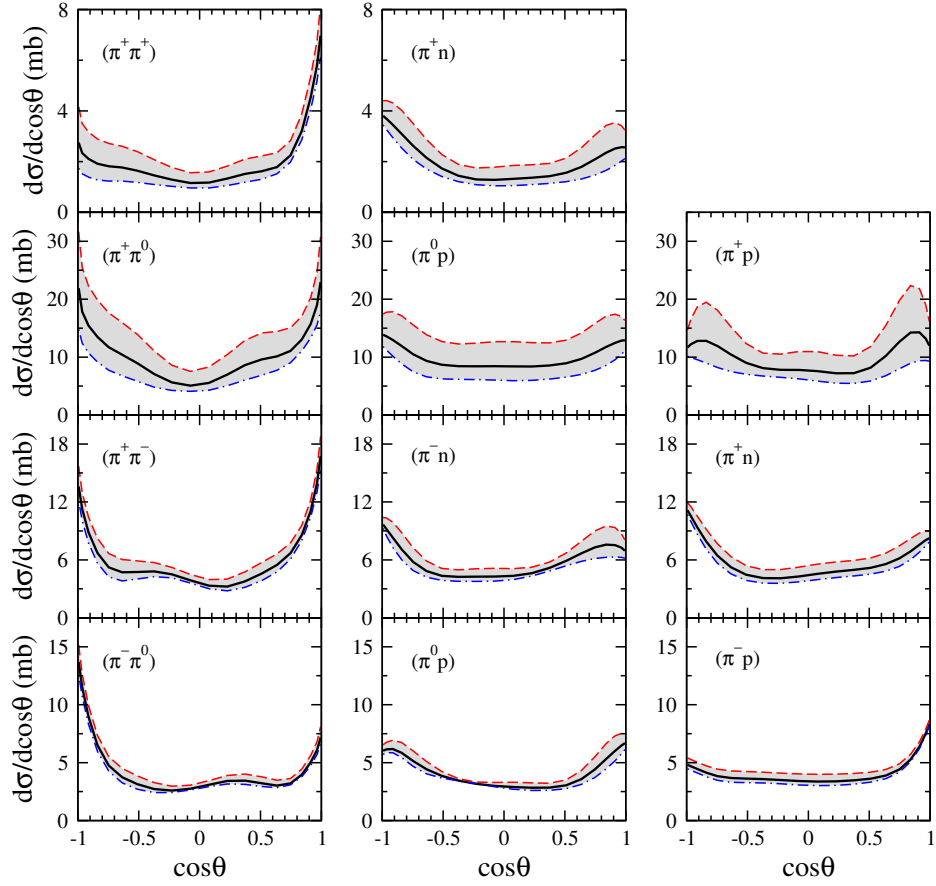}
\caption{
Sensitivity of the angular distributions at $W=1.87$~GeV to 
the $F_{37}[\Delta^*(J^P=\frac{7}{2}^+)] \to \pi\Delta(L=3,S=\frac{3}{2})$ coupling.
First row: $\pi^+ p \to \pi^+ \pi^+ n$;
second row: $\pi^+ p \to \pi^+ \pi^0 p$;
third row: $\pi^- p \to \pi^+ \pi^- n$;
fourth row: $\pi^- p \to \pi^- \pi^0 p$.
The scattering angle $\theta$ denotes the direction of the outgoing two-body subsystem 
indicated in parentheses.
The solid curves denote the full results. 
The grey bands indicate the uncertainty 
resulting from a 50\% variation in the $F_{37} \to \pi \Delta$ vertex magnitude 
(red dashed: $\times 1.5$; blue dash-dotted: $\times 0.5$).
}
\label{fig:F37-piD1-1870-dsdx}
\end{center}
\end{figure}

\begin{figure}[ht]
\begin{center}
\includegraphics[width=0.75\textwidth,angle=0,clip]{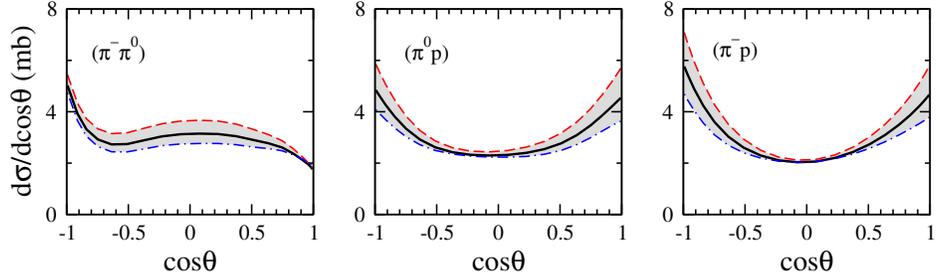}
\caption{
Sensitivity of the angular distributions for $\pi^- p \to \pi^- \pi^0 p$ at $W=1.7$~GeV to 
the $D_{13}[N^*(J^P=\frac{3}{2}^-)] \to \rho N(L=0,S=\frac{3}{2})$ coupling.
The scattering angle $\theta$ denotes the direction of the outgoing two-body subsystem 
indicated in parentheses.
The solid curves denote the full results. 
The grey bands indicate the uncertainty 
resulting from a 50\% variation in the $D_{13} \to \rho N$ vertex magnitude 
(red dashed: $\times 1.5$; blue dash-dotted: $\times 0.5$).
}
\label{fig:D13-rhoN2-1700-dsdx}
\end{center}
\end{figure}

\begin{figure}[ht]
\begin{center}
\includegraphics[width=0.75\textwidth,angle=0,clip]{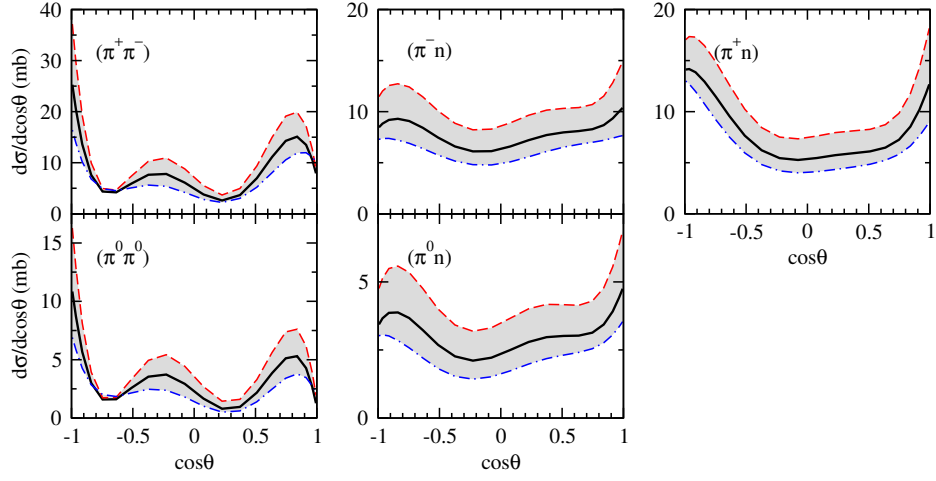}
\caption{
Sensitivity of the angular distributions at $W=1.7$~GeV to 
the $D_{15}[N^*(J^P=\frac{5}{2}^-)] \to \sigma N(L=3,S=\frac{1}{2})$ coupling.
First row: $\pi^- p \to \pi^+ \pi^- n$;
second row: $\pi^- p \to \pi^0 \pi^0 n$.
The scattering angle $\theta$ denotes the direction of the outgoing two-body subsystem 
indicated in parentheses.
The solid curves denote the full results. 
The grey bands indicate the uncertainty 
resulting from a 50\% variation in the $D_{15} \to \sigma N$ vertex magnitude 
(red dashed: $\times 1.5$; blue dash-dotted: $\times 0.5$).
}
\label{fig:D15-sigN1-1700-dsdx}
\end{center}
\end{figure}

\begin{figure}[ht]
\begin{center}
\includegraphics[width=0.75\textwidth,angle=0,clip]{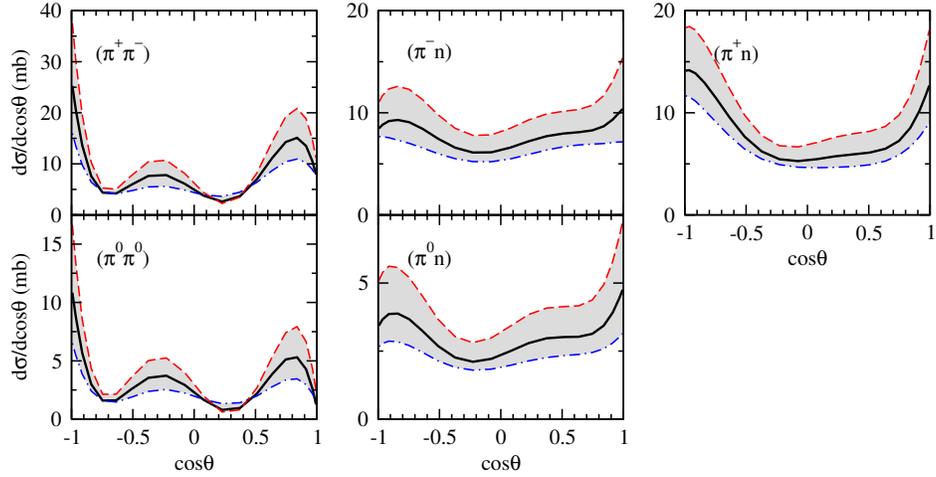}
\caption{
Sensitivity of the angular distributions at $W=1.7$~GeV to 
the $F_{15}[N^*(J^P=\frac{5}{2}^+)] \to \sigma N(L=2,S=\frac{1}{2})$ coupling.
First row: $\pi^- p \to \pi^+ \pi^- n$;
second row: $\pi^- p \to \pi^0 \pi^0 n$.
The scattering angle $\theta$ denotes the direction of the outgoing two-body subsystem 
indicated in parentheses.
The solid curves denote the full results. 
The grey bands indicate the uncertainty 
resulting from a 50\% variation in the $F_{15} \to \sigma N$ vertex magnitude 
(red dashed: $\times 1.5$; blue dash-dotted: $\times 0.5$).
}
\label{fig:F15-sigN1-1700-dsdx}
\end{center}
\end{figure}

In this Appendix, we present the sensitivity of the angular distributions to 
variations in the $\pi\Delta$, $\rho N$, and $\sigma N$ couplings of $N^*$ resonances.
The results are displayed in Figs.~\ref{fig:S31-piD1-1700-dsdx}--\ref{fig:F15-sigN1-1700-dsdx}.
We focus on the same reactions and dressed vertices discussed in Sec.~\ref{sec:sensitivity}, 
where significant effects on the total cross sections and invariant mass distributions were observed.

For invariant mass distributions, the cross section magnitudes are suppressed 
near the kinematic boundaries (maximum and minimum invariant masses) due to phase space effects, 
strictly vanishing at these limits.
Consequently, it is generally difficult to observe sensitivity to 
variations in the $\pi\Delta$, $\rho N$, and $\sigma N$ couplings in these edge regions.
In contrast, angular distributions are not subject to such kinematic suppression 
at forward and backward angles ($\cos\theta \sim \pm 1$); 
thus, the distributions can exhibit significant variations even in these regions
(see, e.g., Figs.~\ref{fig:D15-sigN1-1700-dsdx} and~\ref{fig:F15-sigN1-1700-dsdx}, where
we observe large changes in the magnitude of the distributions for the $\pi\pi$ pair at $\cos\theta \sim -1$).
Furthermore, compared to the invariant mass distributions, 
the angular distributions appear to show more pronounced changes in shape, 
rather than just in overall magnitude (see, e.g., Fig.~\ref{fig:F37-piD1-1870-dsdx}, 
where the $\pm 50\%$ variation of the magnitude of 
$F_{37}[\Delta^*(J^P=\frac{7}{2}^+)] \to \pi\Delta(L=2,S=\frac{3}{2})$ coupling
leads to the appearance or disappearance of a bump structure in the angular distributions
for $\pi^+ p \to \pi^+ \pi^0 p$).
Therefore, to reliably extract $N^*$ information from the $\pi N \to \pi \pi N$ reaction, 
it is highly desirable to have experimental data for both invariant mass and angular distributions.


\end{document}